\documentclass{pas}

\usepackage{multirow}
\usepackage[authoryear]{natbib}
\usepackage{graphicx,amsmath,siunitx,aas-macros}
\usepackage{tikz}
\usepackage{nicefrac}

\lefttitle{Publications of the Astronomical Society of Australia}
\righttitle{Sharma et al.}

\jnlPage{1}{16}
\jnlDoiYr{2026}
\doival{10.1017/pasa.xxxx.xx}

\articletitt{Research Paper}

\title{A benchmark for binary star interaction with a supermassive black hole in general relativity}
\author{
Megha Sharma,$^{1}$\thanks{E-mail: megha.sharma@monash.edu},
Alexander Heger$^{1}$, Daniel J.\ Price$^{1,2}$, Emilio Tejeda$^{3}$, Evgeni Grishin$^{1}$, Luis A.\ Manzaneda$^{4}$, and Alessandro A.\ Trani$^{5,6}$}

\affil{
$^1$School of Physics and Astronomy, Monash University, Wellington Rd, Clayton VIC 3800, Australia \\
$^2$Univ. Grenoble-Alpes, CNRS, Grenoble 38000, France \\
$^3$SECIHTI--Instituto de F\'{i}sica y Matem\'{a}ticas, Universidad Michoacana de San Nicol\'{a}s de Hidalgo, Ciudad Universitaria, 58040 Morelia, Michoac\'{a}n, Mexico \\
$^4$Instituto de Radioastronom\'{i}a y Astrof\'{i}sica, Universidad Nacional Aut\'{o}noma de M\'{e}xico, 58090 Morelia, Michoac\'{a}n, Mexico \\
$^5$Departamento de Astronom\'ia, Facultad Ciencias F\'isicas y Matem\'aticas, Universidad de Concepci\'on, Concepci\'on, 4030000, Chile \\
$^6$National Institute for Nuclear Physics – INFN, Sezione di Trieste, I-34127, Trieste, Italy}

\corresp{M. Sharma, Email: megha.sharma@monash.edu}

\history{(Received xx xx xxxx; revised xx xx xxxx; accepted xx xx xxxx)}

\begin{document}

\begin{abstract}
Most galaxies have supermassive black holes (SMBH) at their centres, surrounded by stars with binary systems also present in this environment. We use two schemes --- post-Newtonian ($\mathrm{PN}$) and a scalar perturbation to a background metric to numerically solve the three-body problem of a binary with a SMBH. We test three different $\mathrm{PN}$ formulations for the $\mathrm{PN}$ scheme: The Einstein-Infeld-Hoffman equation, pair-wise implementation of two-body $\mathrm{PN}$-terms for three bodies and the Arnowitt-Deser-Misner Hamiltonian. We compare these approaches for $10^6\;\mathrm{M}_\odot$ and $10^9\mathrm{M}_\odot$ black holes, and find a statistical match between the two approximations for stellar mass binary interacting with a $10^6\;\mathrm{M}_\odot$ black hole. We also perform a statistical study for encounters with this black hole, and find that the higher order $\mathrm{PN}$ formulation matches with metric-with-perturbation scheme. However, we find a decrease in separation of the binary, and eccentricity variations between different schemes around the $10^9\;\mathrm{M}_\odot$ black hole. This behaviour is not present if binary has a large separation or is further away from the black hole due to decreased general-relativistic effects. We find that the pair-wise $\mathrm{PN}$ method results in a decrease in separation at pericentre in all test cases irrespective of the distance from the black hole or mass of the black hole, making this the least reliable method for solving this problem. Our work highlights the need for caution when interpreting the results in different formulations around SMBHs. This also shows that when understanding extreme mass ratio inspirals (EMRIs) using simulations, one should beware as the binary gets closer to the black hole. 
\end{abstract}
%%%%%%%%%%%%%%%%%%%%%%%%%%%%%%%%%%%%%%%%%%%%%%%%%%

\maketitle
%%%%%%%%%%%%%%%%% BODY OF PAPER %%%%%%%%%%%%%%%%%%

\section{Introduction}
% %EG - I think the first paragraph is too colloquial and redundant.
% The three-body problem has fascinated scientists and the public alike as evident from the success of the book\footnote{\url{https://en.wikipedia.org/wiki/The_Three-Body_Problem_(novel)}} and the Netflix show\footnote{\url{https://en.wikipedia.org/wiki/3_Body_Problem_(TV_series)}} of the same name.  In the show, the alien civilisation struggles to solve this chaotic system for their triple star system, though their problem is exaggerated for dramatic effect. \EG{[[I've already expressed my opinion against opening with this paragraph. The 'real' three body problem is much older and much more fascinating than some corrupt entertainment conglomerates.]]} 
 
The three body problem is ubiquitous in nature and appears in many astrophysical settings. Most stars form in multiple star systems, with the fraction of stars in multiple star systems (multiplicity frequency) increasing from $\sim20\%$ for $M_* < 0.1\;\mathrm{M}_\odot$ to $\sim90\%$ for $M_* > 10\;\mathrm{M}_\odot$ \citep{Offner2023}. Most observed stellar systems are in hierarchical, stable configurations \citep{Perets2026}; Chaotic triples tend to be transient and result in ejections of one of the components \citep[e.g.,][]{stone19, ginat21, kol21}. %But most of these systems are wide and dynamically stable \citep{Perets2026}.

Triple systems with extreme mass ratios are also common: planets and satellites around central stars, as well as stars and stellar remnants around supermassive black holes (SMBH). While there is ample of direct evidence of satellites of giant planets in our own Solar System \citep{nicholsonbook, brozovic17, brozovic22} and formidable literature on their dynamics and long term evolution \citep[e.g.,][]{murraybook, carruba02, cuk04, grishin24b, grishin24a}, the evidence of binary stars and/or compact objects in the Galactic centre environment is only indirect. In particular, photometric observations indicate $\sim 72\%$ of young massive stars in binary configurations at distances beyond the central $1''\ (0.04\ \rm pc)$ for our own Galactic Centre \citep{Gautam2024}. This is in tension with the much lower upper limit of $<47\%$ of the binary fraction within the central $1''$, reported by the same group \citep{Chu23}.

In addition to the dearth of binary stars within the central $1''$, the presence of young S-stars in the Galactic centre allows measuring the SMBH mass of our own galaxy \citep{Ghez1998, Genzel2010, GravityCollab2023}. Most galaxies likely host a central SMBH \citep{Neumayer2020}. The interaction of massive stellar binaries with the central SMBH can lead to a range of outcomes, such as tidal disruption events \citep[TDE,][]{Mandel2015, Bradnick2017} and stellar collisions \citep{Yu2024}. Moreover, \cite{Hills1988} suggested that a tight binary may be disrupted and one of the star may be ejected from the system travelling above the escape speed, called hypervelocity star (HVS). The intense interactions within the central $1''$ may naturally reduce the binary fraction as in the S-star cluster. The surviving binaries require to be tight enough ($<0.1\ \rm au$) to avoid disruption from the SMBH and passing stars \citep{Stephan2019, Dodici2025}.

%\EG{TBD [A a paragraph that delineates the necessity of accurate modelling of the binary + smbh system: importance of GR and various approaches, and why a code comparison/benchmark paper like this work is important.] }

%In this work we consider the three-body problem, but instead of a triple-star system, we consider a stellar binary with a supermassive black hole (SMBH). 

%Almost all galaxies harbour SMBHs at their centre surrounded by stars \citep{Neumayer2020}. Our Galactic Centre also has a SMBH at its centre \citep{Ghez1998,Genzel2010,GravityCollab2023}. Observations suggest that $\sim72\%$ ($68\%$ confidence) of stars in the Galactic Centre belong to young, massive star binary fraction in the central $\sim0.4\,\mathrm{pc}$, consistent with the local solar neighbourhood binary fractions of the OB stars \citep{Gautam2024}. These binary stars could interact with the SMBH, resulting in the need to understand the general relativistic (GR) three-body problem. \citet{Stephan2019,Dodici2025} argued that most of these binaries would be tight (semi-major axis $< 0.1\;\mathrm{au}$) due to resonant relaxation and perturbations from the passing stars. 

The close encounter of a binary in the vicinity of a SMBH requires delicate treatment that accurately incorporates the effects of general relativity (GR). Previous studies have focused mainly on lower mass SMBHs with weaker GR effects and used Newtonian gravity to investigate the problem \citep[e.g.,][]{Gould2003, Gualandris2005, Sari2010, Sersante2025}. \citet{Mandel2015} found that a binary star system approaching a SMBH can result in double tidal disruption events. \citet{Yu2024} also used Newtonian physics to determine rates of collision of binary stars around $10^6\;\mathrm{M}_\odot$ SMBH. 

Recently, \citet{Manzaneda2024} used a Hybrid-Relativistic-Newtonian-Approximation (\textsc{Hrna}) combining the Schwarzschild metric of the SMBH and the Newtonian self-gravity of the binary to explore hypervelocity star formation, and argued that the stars could have faster velocities in the general relativistic (GR) regime than in Newtonian physics, which could result in exotic events. \citet{Manzaneda2024} found that the number of collision products is higher for simulations in \textsc{Hrna} compared with the Newtonian physics. All these simulations considered a $10^6\;\mathrm{M}_\odot$ black hole.

%The three-body problem of a binary star with a black hole has been explored before. \citet{Hills1988} proposed that the interaction of a binary star system with a SMBH can result in one star being ejected while other ending bound to the black hole. The escaping star is termed as a hypervelocity star due to its high velocity. \citet{Gould2003,Gualandris2005,Sari2010,Sersante2025} have used Newtonian gravity to understand this phenomenon. \citet{Manzaneda2024} used a Hybrid-Relativistic-Newtonian-Approximation (\textsc{Hrna}) combining the Schwarzschild metric of the SMBH and the Newtonian self-gravity of the binary to explore hypervelocity star formation, and argued that the stars could have faster velocities in general relativistic (GR) regime than in Newtonian physics.
%Using Newtonian physics, \citet{Mandel2015} found that a binary star system approaching a SMBH can result in double tidal disruption events. \citet{Yu2024} also used Newtonian physics to determine collision of binary stars around $10^6\;\mathrm{M}_\odot$ SMBH. They argued that the collision products of such events would result in exotic events. \citet{Manzaneda2024} found that the number of collision products is higher for simulations in \textsc{Hrna} compared with the Newtonian physics. All these simulations considered a $10^6\;\mathrm{M}_\odot$ black hole.

The correct treatment of the problem where a binary approaches a SMBH would require numerical relativity, which is numerically expensive and difficult. Instead of using Newtonian gravity, several methods can be used to model the system in GR. The accuracy of commonly used approximation methods for three-body problem remains unclear. These approaches include --- post-Newtonian approximations ($\mathrm{PN}$) \citep{Galaviz2011,Trani2023} and metric-with-perturbation techniques such as \textsc{Hrna} \citep{Manzaneda2024}.  

Generally, the $\mathrm{PN}$ approach includes adding terms of the order $v/c$ to the Newtonian potential \citep{Einstein1938,Bahcall1976,Iyer1993}. The Einstein-Infeld-Hoffman (\textsc{EIH}) equation \citep{Einstein1938} describes the $1\mathrm{PN}$ acceleration for N-bodies. The higher order $\mathrm{PN}$ terms were first derived for binary systems \citep{Blanchet1998}, with current knowledge of up to $4.5\mathrm{PN}$ terms \citep{Blanchet2024}. 

Codes such as \textsc{Tsunami} \citep{Trani2023} use post-Newtonian corrections implemented in a pair-wise manner for a triple system. \textsc{EIH} was simplified for a $N$-body system around a SMBH by \citet{Will2014}, ignoring the star-star interactions to derive $\mathrm{PN}$ approximation equations. The key difference between the pair-wise \textsc{PN} and \textsc{EIH} is that pair-wise methods ignore the cross-terms \citep{Will2014, Zwart2022}. Another $\mathrm{PN}$ approach is to solve the Arnowitt-Deser-Misner (\textsc{ADM}) Hamiltonian \citep{Arnowitt1959} that has been derived for a triple system up to $2.5\mathrm{PN}$ \citep{Schafer1987,Lousto2008,Galaviz2011}. But the \textsc{EIH} equations are written in harmonic coordinates while the \textsc{ADM} Hamiltonian uses canonical coordinates. The equivalence between the two approaches has been determined for a binary system by \citet{Damour2001} and \citet{Blanchet2004}, but not for the three-body system. The post-Newtonian approximation is only accurate when $v/c \ll 1$, which breaks down close to a SMBH. Recently, the $2\mathrm{PN}$ \textsc{ADM} formalism has been extended to $N$-body system \citep{Heinze2026}. 

In this work we compare different techniques to solve the three-body general relativistic problem involving a stellar mass binary plunging towards a central SMBH, using two different central black hole masses, $10^6\;\mathrm{M}_\odot$ and $10^9\;\mathrm{M}_\odot$. We use our in-house $\mathrm{PN}$ code, \textsc{Multistar} which uses pair-wise $\mathrm{PN}$ corrections up to $3.5\mathrm{PN}$, and also solve the equations of motion from the \textsc{ADM} Hamiltonian. We also solve the \textsc{EIH} equations for comparison. We have expanded \textsc{Geodesic} \citep{Liptai2019}, similar to \citet{Manzaneda2024} to include a perturbation to the metric. We call the new version of \textsc{Geodesic} as \textsc{Phantom-Geo}.

Our goal is to understand which method is most suitable for this problem. Our paper is structured as follows: Section~\ref{sec:methods} describes our method. Section~\ref{sec:results} lists the results and we discuss them in Section~\ref{sec:discussion}. We conclude in Section~\ref{sec:conclusion}.

\section{Methods}
\label{sec:methods}
We now describe the post-Newtonian and metric-with-\\perturbation codes. We evolve the same initial conditions in all codes for comparison. All methods assume point-mass particles. 

\subsection{Codes with post-Newtonian approximation}
\subsubsection{Einstein-Infeld-Hoffman (\textsc{EIH}) equation}
\label{sec:EIH}
The \textsc{EIH} equations were first derived by \citet*{Einstein1938,Infeld1957}, and it describes the $1\mathrm{PN}$ interaction of $N$-body point-mass systems. It is written in harmonic coordinates which satisfies the harmonic gauge ($\nabla^{\mu} \nabla_{\mu} x^\lambda = 0 = g^{\rho \sigma} \Gamma^\lambda_{\rho \sigma}$, where $x^\lambda$ is the spacetime coordinate, and $\Gamma$ is Christoffel symbol; \citealt{Carroll1997}). 
The three-body problem has not yet been expanded to higher order $\mathrm{PN}$ terms in this coordinate system.  The acceleration of each body is given by
\begin{align}
\mathbf{a}_i &= 
 -\sum_{j\ne i} \frac{G m_j\, \mathbf{r}_{ij}}{r_{ij}^3}
+ \frac{1}{c^2} \sum_{j\ne i} 
\frac{G m_j\, \mathbf{r}_{ij}}{r_{ij}^3}
\Bigg[
4\,\frac{G m_j}{r_{ij}}
+ 5\,\frac{G m_i}{r_{ij}} \nonumber\\
&\quad+ \sum_{k\ne i,j} \frac{G m_k}{r_{jk}}
+ 4 \sum_{k\ne i,j} \frac{G m_k}{r_{ik}}
- \frac{1}{2} \sum_{k\ne i,j} 
\frac{G m_k}{r_{jk}^3} (\mathbf{r}_{ij}\!\cdot\!\mathbf{r}_{jk}) \nonumber\\
&\quad- v_i^2 + 4\,\mathbf{v}_i\!\cdot\!\mathbf{v}_j
- 2 v_j^2
+ \frac{3}{2}(\mathbf{v}_j\!\cdot\!\mathbf{n}_{ij})^2
\Bigg] \nonumber\\
&\quad - \frac{7}{2c^2}
\sum_{j\ne i} \frac{G m_j}{r_{ij}}
\sum_{k\ne i,j} \frac{G m_k\, \mathbf{r}_{jk}}{r_{jk}^3} \nonumber\\
&\quad + \frac{1}{c^2}\sum_{j\ne i} \frac{G m_j}{r_{ij}^3}
\mathbf{r}_{ij}\cdot(4\mathbf{v}_i - 3\mathbf{v}_j)\mathbf{v}_{ij}\;,
\label{eq:EIH}
\end{align}
where $m$ is mass, $\mathbf{r}$ is position vector, $\mathbf{v}$ is the velocity vector, $\mathbf{r}_{ij} = \mathbf{r}_i-\mathbf{r}_j$, and $\mathbf{n}_{ij} = \mathbf{r}_{ij}/r_{ij}$. 
We solve this using the $4$th order Runge-Kutta method \citep{Butcher1996}. 

\subsubsection{Acceleration equations derived from \textsc{EIH} equation}

\citet{Will2014} derived the $\mathrm{PN}$ accelerations directly from the \textsc{EIH} equations for $N$ bodies around a SMBH (Section~III in \citealp{Will2014}). The acceleration equations of the stars and the SMBH are written up to the $G^2m^2/c^2r^3$ order, ignoring the star-star interaction. The acceleration of the black hole (index $1$) is given by 
\begin{align}
\mathbf{a}_1 &= -\sum_j \frac{G m_j \mathbf{r}_{1j}}{r_{1j}^3}
+ \frac{1}{c^2} \mathbf{a}_{\text{BH}}
+ \frac{1}{c^2} \mathbf{a}_{\text{cross}}
+ O\!\left( \frac{G^2 m_j^3}{M c^2 r^3} \right)\;,
\end{align}
where 
\begin{align}
\mathbf{a}_{\text{BH}}
&= \sum_j \frac{G m_j \mathbf{r}_{1j}}{r_{1j}^3}
\left( \frac{5Gm_1}{r_{1j}} - 2 v_j^2 + \frac{3}{2} (\mathbf{v}_j \cdot \mathbf{n}_{1j})^2 \right)\nonumber\\
&\quad+ 3 \sum_j \frac{G m_j}{r_{1j}^3} (\mathbf{v}_j \cdot \mathbf{r}_{1j}) \mathbf{v}_j\;,
\end{align}
and 
\begin{align}
\mathbf{a}_{\text{cross}} &=
4 \sum_j \frac{G^2 m_j^2 \mathbf{r}_{1j}}{r_{1j}^4} 
+ \sum_{j,k} \frac{G^2 m_j m_k \mathbf{r}_{1j}}{r_{1j}^3}
\left( \frac{4}{r_{1k}} + \frac{5}{4 r_{jk}} - \right. \nonumber \\ 
&\quad \left. \frac{r_{1k}^2}{4 r_{jk}^3}
+ \frac{r_{1j}^2}{4 r_{jk}^3} \right)
- \frac{7}{2} \sum_{j,k} \frac{G^2 m_j m_k \mathbf{r}_{jk}}{r_{jk}^3 r_{1j}}
- \sum_{j,k} \frac{G m_j m_k}{M r_{1j}^3} \nonumber\\
&\quad
\left[ 4 (\mathbf{v}_j \cdot \mathbf{v}_k) \mathbf{r}_{1j}
- 3 (\mathbf{v}_j \cdot \mathbf{r}_{1j}) \mathbf{v}_k
- 4 (\mathbf{v}_k \cdot \mathbf{r}_{1j}) \mathbf{v}_j \right]\;,
\end{align}
are the $1\mathrm{PN}$ term, with $j$ and $k$ corresponding to the stars index. The acceleration of the stars is given by
\begin{align}
\mathbf{a}_i
&= -\frac{Gm_1\, \mathbf{r}_{i1}}{r_{i1}^3}
- \sum_j \frac{G m_j \mathbf{r}_{ij}}{r_{ij}^3}
+ \frac{1}{c^2}\mathbf{a}_{i,\mathrm{BH}}
+ \frac{1}{c^2}\mathbf{a}_{i,\mathrm{cross}} \nonumber\\
&\quad+ O\!\left( \frac{G^2 m_j^2}{c^2 r^3} \right)\;,
\end{align}
where $i$ and $j$ are index of stars and $1$ corresponds to the SMBH. $\mathbf{a}_{i,\mathrm{BH}}$ and $\mathbf{a}_{i,\mathrm{cross}}$ are given by
\begin{align}
\mathbf{a}_{i,\mathrm{BH}}
&=
\frac{Gm_1\, \mathbf{r}_{i1}}{r_{i1}^3}
\left( \frac{4Gm_1}{r_{i1}} - v_i^2 \right)
+ 4 \frac{G m_1}{r_{i1}^3} (\mathbf{v}_i \cdot \mathbf{r}_{i1})\mathbf{v}_i\;,
\end{align}
and
\begin{align}
\mathbf{a}_{i,\mathrm{cross}}
&=
5 \frac{G^2 m_i m_1\, \mathbf{r}_{i1}}{r_{i1}^4}
- \frac{G m_i}{r_{i1}^3}
\left[ 4 v_i^2 \mathbf{r}_{i1}
- 7 (\mathbf{v}_i \cdot \mathbf{r}_{i1}) \mathbf{v}_i \right]
\nonumber \\
&\quad
+ \sum_j \frac{G^2 m_j m_1\, \mathbf{r}_{i1}}{r_{i1}^3}
\left( \frac{4}{r_{ij}}
+ \frac{5}{4 r_{j1}}
+ \frac{r_{i1}^2}{4 r_{j1}^3}
- \frac{r_{ij}^2}{4 r_{j1}^3} \right)
\nonumber \\
&\quad
+ \sum_j \frac{G^2 m_j m_1\, \mathbf{r}_{ij}}{r_{ij}^3}
\left( \frac{4}{r_{i1}}
+ \frac{5}{4 r_{j1}}
- \frac{r_{i1}^2}{4 r_{j1}^3}
+ \frac{r_{ij}^2}{4 r_{j1}^3} \right)
\nonumber \\
&\quad
- \frac{7}{2} \sum_j \frac{G^2 m_j m_1\, \mathbf{r}_{j1}}{r_{j1}^3}
\left( \frac{1}{r_{ij}} - \frac{1}{r_{i1}} \right)
\nonumber \\
&\quad
- \sum_j \frac{G m_j}{r_{i1}^3}
\left[ 4 (\mathbf{v}_i \cdot \mathbf{v}_j)\mathbf{r}_{i1}
- 3 (\mathbf{v}_j \cdot \mathbf{r}_{i1}) \mathbf{v}_i
- 4 (\mathbf{v}_i \cdot \mathbf{r}_{i1})\mathbf{v}_j \right]
\nonumber \\
&\quad
+ \sum_j \frac{G m_j \mathbf{r}_{ij}}{r_{ij}^3}
\left[ v_i^2 - 2 v_{ij}^2
+ \frac{3}{2} (\mathbf{v}_j \cdot \hat{\mathbf{r}}_{ij})^2 \right]
\nonumber \\
&\quad
+ \sum_j \frac{G m_j}{r_{ij}^3}
\left[ \mathbf{r}_{ij} \cdot (4\mathbf{v}_i - 3 \mathbf{v}_j) \right]
\mathbf{v}_{ij}\;,
\end{align}
respectively. This formulation of \textsc{EIH} reduces the numerical complexity from $O(n^3)$ to $O(n^2)$. We again use the $4$th order Runge-Kutta method for the integration of these equations. From here on, we would refer to the results of this method as \textsc{Will14}.

\subsubsection{Pair-wise $\mathrm{PN}$ implementation}
We use a pair-wise post-Newtonian code, \textsc{Multistar} with up to $3.5\mathrm{PN}$ terms. \textsc{Multistar} uses hierarchical or Jacobi coordinates, where coordinates are defined such that we consider the relative motion between the binary star, and the motion of SMBH relative to the centre of mass of the binary (similar to \citealt{Mardling2002} for the four body problem). The positions and velocities are integrated in this coordinate frame. We assume all bodies are non-spinning.

To determine the forces, we convert Jacobi coordinates to relative vectors and the forces are calculated pair-wise. The Newtonian gravitational force on object $i$ from the object $j$ is given by
\begin{equation}
\mathbf{F}_{ij,\mathrm{N}} = -\frac{G m_i m_j}{r_{ij}^2} \mathbf{n}_{ij}\;.
\end{equation} 
We next implement pair-wise post-Newtonian terms. The equation of motion is conserved for even $\mathrm{PN}$ terms, where $1\mathrm{PN}$ introduces precession in the orbit \citep{Einstein1938,Merritt2013}, $2\mathrm{PN}$ and $3\mathrm{PN}$ introduce energy terms which refine the precession rate \citep{Blanchet2014}. Odd $\mathrm{PN}$ terms result in gravitational wave emission from the orbit \citep{Blanchet1998, Kidder1995,Blanchet2014}. The $1\mathrm{PN}$ is given by \citet{Einstein1938,Damour1985}
\begin{align}
\mathbf{F}_{ij,1\mathrm{PN}} &= -\frac{G m_i m_j}{r_{ij}^2 c^2 } \left\{ \left[\left (1 + 3 \eta\right) v_{ij}^2
  - 2\left(2+\eta\right) \frac{G m_{ij}}{r_{ij}} - \frac{3}{2} \eta \dot{r}_{ij}^2 \right] \mathbf{n}_{ij} \right. \nonumber \\
&\quad \left. - 2\left(2 - \eta\right)\dot{r}_{ij} \textbf{v}_{ij} \right\}\;,
\end{align}
where $\eta = m_i m_j/m_{ij}^2$, $m_{ij} = m_i + m_j$ and $\dot{r}_{ij} = \textbf{n}_{ij} \cdot \textbf{v}_{ij} $.
The $2\mathrm{PN}$ term \citep{Blanchet1998} is 
\begin{align}
\label{eq:2pn}
\mathbf{F}_{ij,2\mathrm{PN}} &= -\frac{G m_i m_j}{r_{ij}^2c^4} \left\{ \left[
\eta \dot{r}_{ij}^4 \left( \frac{15}{8} - \frac{45\eta}{8} \right) 
+ \eta v_{ij}^4 \left( 3 - 4\eta \right) \right.\right.\nonumber\\
&\quad \left. \left. 
+ \eta\left(-\frac{9}{2} + 6  \right) \dot{r}_{ij}^2v_{ij}^2 
-\frac{G m_{ij}\dot{r}_{ij}^2}{r_{ij}} \left(2 + \eta \left( 25 + 2 \eta v_{ij}^2\right)\right)
\right. \right. \nonumber \\
&\quad \left. \left.
- \frac{G m_{ij}v_{ij}^2}{r_{ij}} \left( \eta \left( \frac{13}{2} - 2 \eta\right)\right) 
+ \frac{G^2 m_{ij}^2}{r_{ij}^2} \left(9 + \frac{87\eta}{4}\right)
\right] \textbf{n}_{ij} \right. \nonumber \\
&\quad \left.
 + \left[ -\eta v_{ij}^2 \left(\frac{15}{2}+ 2\eta \right)
+ \eta \left( \frac{9}{2} + 3\eta\right) \dot{r}_{ij}^2 
 \right. \right. \nonumber \\
 &\quad \left. \left.
+ \frac{G m_{ij}\eta }{r_{ij}} \left( 2 +\eta \left(\frac{41}{2} + 4 \eta \right) \right)\right] \dot{r}_{ij} \mathbf{v}_{ij}
\right\}\;.
\end{align}
The $2.5\mathrm{PN}$ term \citep{Blanchet1998} for relativistic radiation (gravitational waves) is 
\begin{align}
    \mathbf{F}_{ij,2.5\mathrm{PN}} &= \frac{8 G m_i m_j m_{ij} \eta}{5 r_{ij}^3 c^5}
    \left[ 
    \left(\frac{17 G m_{ij}}{3r_{ij}} + 3 v_{ij}^2\right)\dot{r_{ij}}\,\textbf{n}_{ij}
    \right. \nonumber \\
    &\quad \left.
    - \left( \frac{3 G m_{ij}}{r_{ij}} + v_{ij}^2 \right) \textbf{v}_{ij}
    \right]\;.
\end{align}
The $3\mathrm{PN}$ and $3.5\mathrm{PN}$ terms are listed in \ref{app:post_new_terms}.  We alternatively tested the $2.5\mathrm{PN}$ term from \citet{Kidder1995}
\begin{align}
    \mathbf{F}_{ij,2.5\mathrm{PN}} &= \frac{8 G m_i m_j m_{ij} \eta}{5 r_{ij}^3 c^5}
    \left[ 
    \left(\frac{2 G m_{ij}}{3r_{ij}} + 18 v_{ij}^2 - 25\dot{r}_{ij}^2 \right)\dot{r}_{ij}\,\textbf{n}_{ij}
    \right. \nonumber \\
    &\quad \left.
    + \left( \frac{2 G m_{ij}}{r_{ij}} - 6v_{ij}^2 +15\dot{r}_{ij}^2\right) \textbf{v}_{ij}
    \right]\;,
\end{align}
and found no significant differences in our results. The results shown in this paper use the \citet{Blanchet1998} term. We use point particles as our stars and black hole and use Bulirsch–Stoer integrator. The method determines the solution at a step-size of `$h$' where $h$ can be large. The sub-steps are performed using the mid-point method and Richardson extrapolation is used to determine the solution as if the step size goes to zero \citep{Bulirsch1966}. For each $\mathrm{PN}$ order, we include the Newtonian term and all $\mathrm{PN}$ corrections up to that order.

In addition, we employ the $\mathrm{PN}$ code \textsc{Tsunami} \citep{Trani2023}. \textsc{Tsunami} uses chain coordinates such that a relative vector is defined for the closest objects, followed by the third object with the closest object of the binary. The positions and velocities are integrated in this coordinate system, resulting in one less equation of motion. It also reduces the round-off errors \citep{Mikkola1993}. The $\mathrm{PN}$ terms are implemented pair-wise similar to \textsc{Multistar} but it uses accelerations instead of forces.
\textsc{Tsunami} uses regularisation of the equation of motion which deals with the singularity. It also uses Bulirsch–Stoer extrapolation for integration like \textsc{Multistar}. We compare both codes for up to $3.5\mathrm{PN}$.
\begin{figure*}
    \centering
    \includegraphics[width=0.9\textwidth,height=0.9\textheight, keepaspectratio]{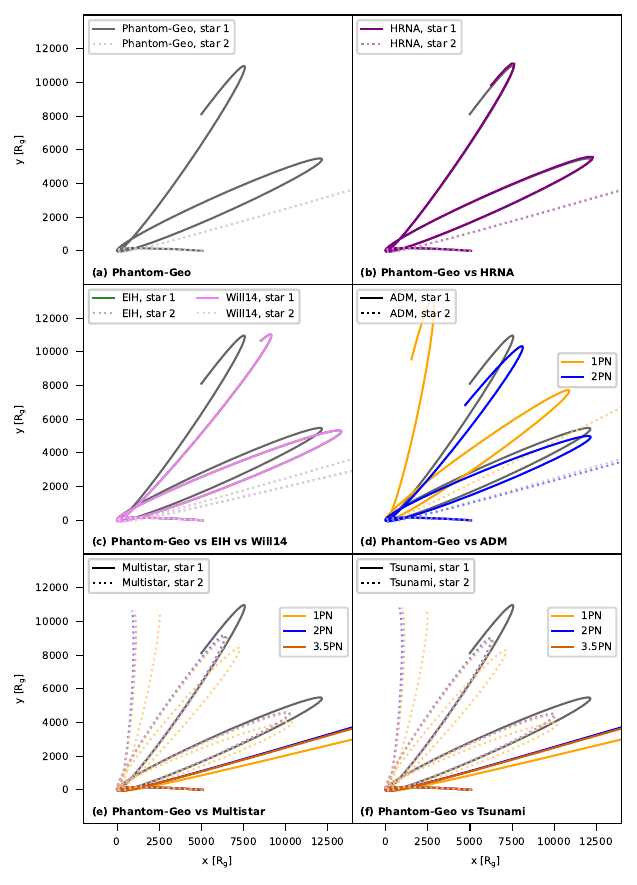}
    \caption{Orbits of $0.5\;\mathrm{M}_\odot$ binary stars around a $10^6\;\mathrm{M}_\odot$ SMBH. This model was obtained from \citet{Manzaneda2024}. We zoom onto the bound star to compare precession. (a) In \textsc{Phantom-Geo}, star $1$ gets bound to the black hole while star $2$ escapes. (b) \textsc{Phantom-Geo} vs \textsc{Hrna}, both codes match with each other. (c) \textsc{Phantom-Geo} comparison with \textsc{EIH} (green lines), and \textsc{Will14} (pink lines). \textsc{EIH} and \textsc{Will14} matches with each other. (d) \textsc{Phantom-Geo} vs \textsc{ADM}. yellow and blue lines correspond to $1\mathrm{PN}$ and $2\mathrm{PN}$, respectively. The amplitude of precession are similar, but the rates are different. (e) \textsc{Phantom-Geo} vs \textsc{Multistar}, where opposite stars get bound and unbound. (f) \textsc{Phantom-Geo} vs \textsc{Multistar} vs \textsc{Tsunami}, where both pair-wise $\mathrm{PN}$ codes match with each-other.}
    \label{fig:figure1}
\end{figure*}

% \begin{figure*}
%     \centering
%       \includegraphics[width=0.9\textwidth,height=0.9\textheight,keepaspectratio]{fig1_full2.pdf}
%     \caption{Orbits of $0.5\;\mathrm{M}_\odot$ stars around a $10^6\;\mathrm{M}_\odot$ black hole in harmonic frame except for \textsc{ADM} for which we do not perform a coordinate transformation. \textsc{EIH} and \textsc{Will14} agree with each other. $2\mathrm{PN}$ \textsc{ADM} is close to \textsc{EIH}. $3.5\mathrm{PN}$ \textsc{Multistar} binary evolution is closer to \textsc{Phantom-Geo} compared with lower $\mathrm{PN}$ values.  }
%     \label{fig:figure_sep_rad_test2}
% \end{figure*}
% \begin{figure*}
%     \centering
%     \includegraphics[width=\textwidth,height=\textheight, keepaspectratio]{fig1_full_sch_panels_test2.pdf}
%     \caption{Separation and eccentricity vs time for model shown in Figure~\ref{fig:figure_sep_rad_test2}. \textsc{Multistar} and \textsc{Tsunami} match with each other. Binary becomes eccentric post-pericentre for the \textsc{EIH} and \textsc{Will14} codes. An increase to upto eccentricity of $0.2$ is also noticed for \textsc{ADM} codes. }
%     \label{fig:figure1_panels_test2}
% \end{figure*}
\subsubsection{\textsc{ADM} Hamiltonian approach}
The Arnowitt-Deser-Misner (\textsc{ADM}; \citealt*{Arnowitt1959,Arnowitt2008}) approach describes the Hamiltonian where the dynamics are described in a three-dimensional hypersurface where the fields are defined \citep{Corichi2022}. In this formalism, the spacetime is split into (3+1) coordinates of space and time \citep{Schafer2018}. For triple systems, the \textsc{ADM} canonical Hamiltonian has been written up to $2.5\mathrm{PN}$ \citep{Schafer1987,Lousto2008,Galaviz2011}, but corresponding harmonic coordinate equations of motion have not been derived yet.  The Newtonian Hamiltonian is given by
\begin{align}
H_0 = \frac{1}{2} \sum_{i} \frac{\mathbf{p}_i^{\,2}}{m_i}
      - \frac{1}{2} \sum_{i} \sum_{j\ne i} \frac{G m_i m_j}{r_{ij}} \, ,
\end{align}
where $\mathbf{p}_i = m_i \mathbf{v}_i$. The $1\mathrm{PN}$ Hamiltonian is given by
\begin{align}
H_1 &= 
-\frac{1}{8} \sum_{i} m_i \left( \frac{\mathbf{p}_i^{\,2}}{m_i^2} \right)^{2}
-\frac{1}{4} \sum_{i} \sum_{j\ne i} \frac{G}{r_{ij}}
\left(
    6\,\frac{m_j}{m_i}\,\mathbf{p}_i^{\,2}
    - 7\,\mathbf{p}_i \!\cdot\! \mathbf{p}_j \right. \nonumber\\
&\quad \left.
    - (\mathbf{n}_{ij} \!\cdot\! \mathbf{p}_i)(\mathbf{n}_{ij} \!\cdot\! \mathbf{p}_j)
\right)
+ \frac{G^2}{2} \sum_{i} \sum_{j\ne i} \sum_{k\ne i}
\frac{m_i m_j m_k}{r_{ij} r_{ik}} \, .
\end{align}
The $2\mathrm{PN}$ Hamiltonian ($H_2$) is described in \ref{app:ADM2pn}. We found a typo in \citet{Galaviz2011}, which was not present in \citet{Lousto2008} ($r_\mathrm{ij}^2$ -> $r_\mathrm{ij}$ term, colored red in Equation~\ref{eq:adm2pn_eq}). The total Hamiltonian is
\begin{align}
    H &= H_0 + \frac{1}{c^2} H_1 + \frac{1}{c^4} H_2\;.
\end{align}
We derive the equations of motion using \textsc{sympy} \citep{sympy}, and use Runge-Kutta method to evolve the system. 

We now turn to the metric-with-perturbation approach, where the binary is modelled as a perturbation to a fixed relativistic background.

\subsection{Codes with binary stars as a perturbation to the metric}
We use the general relativistic code \textsc{Geodesic}, \citep{Liptai2019}, and add the binary as a Newtonian perturbation to the metric\footnote{We use G=c=1, and black hole mass, $M = M_\bullet/10^6$ in the code.}. We only consider point mass particles, which implies that the pressure ($P$) and internal energy ($u$) are $0$ in relativistic hydrodynamics equations, resulting in enthalpy ($w$) of $1$.  Hence, we can write the conserved momentum and energy \citep{Monaghan2001} as
\begin{subequations}
\begin{align}
    \label{eq:cons_terms}
    p_i &= \frac{U^0 v_i}{c^2}\;,\\
    e &=  \frac{[1 + v_iv^i]}{U^0c^2}\;,
\end{align}
\end{subequations}
where $v^\mu = dx^\mu/dt$,  $U^0 = dt/d\tau = 1/\sqrt{-\tilde{g}_{\mu\nu}v^\mu v^\nu / c^2}$, where $\tau$ is the proper time and $t$ is an Eulerian observer's time, and
\begin{align}
\label{eq:metric}
    \tilde{g}_{\mu\nu} &= g_{\mu\nu} + h_{\mu\nu}\;,
\end{align}
where $ g_{\mu\nu}$ is the Kerr or Schwarzschild metric (we use sign of (-, +, +, +) for the metric) and $h_{\mu\nu}$ is the Newtonian perturbation to the metric given by \citep{Bardeen1980,Mukhanov1992}
\begin{subequations}
\begin{align}
\label{eq:poten1}
    h_{00} &= -2 \Phi/c^2\;, \\
    \label{eq:poten2}
    h_{ij} &= -2 \Phi \delta_{ij}/c^2\;,
\end{align}
\end{subequations}
where $\Phi$ is Newtonian gravitational potential between stars ($\Phi = - G m / r$, where $m$ is mass of the star and $r$ is the distance between the stars). In the rest frame of the fluid, the conserved quantities can be written as
\begin{subequations}
\begin{align}
    p_i &= \frac{\Gamma V_i}{c^2}\;,\\
    e &= p_iv^i + \frac{\alpha c^2}{\Gamma}\;,
\end{align}
\end{subequations}where $\Gamma = (1 - V^i V_i/c^2)^{-1/2}$ is the generalised Lorentz factor. $V^i$ is the fluid velocity in the frame of local Eulerian observer given by $(v^i + \beta^i)/\alpha $. $\alpha$, $\beta$ and $\gamma$ are related to the four metric \citep{Liptai2019}. 

The change in momentum for each point mass is given by
\begin{align}
\frac{dp_i}{dt} &= \frac{1}{2}U^0 v^\mu v^\nu \frac{\partial \tilde{g}_{\mu\nu}}{\partial x^i}\;.
\label{eq:momentum_main}
\end{align}
Using Equations~(\ref{eq:metric})-(\ref{eq:poten2}), we can rewrite Equation~(\ref{eq:momentum_main}) as 
\begin{subequations}
\begin{align}
    \frac{dp_i}{dt} &= \frac{1}{2} U^0 v^\mu v^\nu
    \frac{\partial g_{\mu  \nu}}{\partial x^i} - U^0 \frac{\partial \Phi}{\partial x^i} - U^0 \delta_{lm} \frac{v^l v^m}{c^2} \frac{\partial\Phi}{\partial x^i}\;.
\end{align}
\end{subequations}
The 2nd term is the Newtonian acceleration, while 3rd term corresponds to special relativistic correction of order $(v/c)^2$. Since the GR effects are dominant only close to the black hole, we ignore the third term. We also assume $U^0 = 1$ in the 2nd term which means that the observer's time is same as proper time, which is true far from the black hole, where the binary interaction is more important. Thus, the final form of our momentum equation is 
\begin{align}
\label{eq:mom_eq}
    \frac{dp_i}{dt} &= \frac{1}{2} U^0 v^\mu v^\nu
    \frac{\partial g_{\mu  \nu}}{\partial x^i} - \frac{\partial \Phi}{\partial x^i}\;,
\end{align}
where we keep the $U^0$ in the first term. We evolve the momentum and positions. First, we convert the initial velocities to momentum using Equation (\ref{eq:cons_terms}), and then use the hybrid leapfrog method (\citealt{Leimkuhler2005}, see Equations ($79$)-($83$) in \citealt{Liptai2019}).  We calculate the velocities from the momenta at different times using 
\begin{equation}
    v^i  = \tilde{g}^{ij} \left(\frac{\alpha p_j}{\Gamma} - \beta_j\right)\;.
\end{equation}
\ref{app:geodesic_test} shows a test of a isolated binary star system. We evolve the model for $100$ orbits, and find that the angular momentum and energy are conserved to a typical relative error of $10^{-11}$ (see Figure~\ref{fig:binary_star_energy}). We name this new version of \textsc{Geodesic} as \textsc{Phantom-Geo}. 

This approach is slightly different from \textsc{Hrna} \citep{Manzaneda2024}. Unlike us, they evolve acceleration and positions of two point particles. Using the geodesic equation, and a four acceleration, $a^\alpha$, external force of a point mass can be written as
\begin{equation}
\label{eq:geodesic}
    \frac{d^2x^\alpha}{d\tau^2} + \Gamma^\alpha_{\mu\nu} \frac{dx^\mu}{d\tau}\frac{dx^\nu}{d\tau} =  a^\alpha \;,
\end{equation}
where $\Gamma^\alpha_{\mu\nu}$ are the Christoffel symbols, and 
\begin{equation}
    a^\alpha = P^{\alpha\beta} \frac{\partial \Phi}{\partial x^\beta}=(g^{\alpha\beta} + U^\alpha U^\beta) \frac{\partial \Phi}{\partial x^\beta} \;.
\end{equation}
The Newtonian gravity is corrected using a projection tensor $P^{\alpha\beta}$ which ensures that the four-acceleration is always orthogonal to the four velocity. Using the time component of Equation (\ref{eq:geodesic}), the remaining spatial components can be written as \citep{Tajeda2017}
\begin{equation}
\label{eq:hmra_vel}
    \frac{dv^i}{dt} = (-g^{i\lambda} + \frac{v^i}{c} g^{0\lambda}) \left[ \frac{1}{(U^0)^2} \frac{\partial \Phi}{\partial x^\lambda} + \left(\frac{\partial g_{\mu\lambda}}{\partial x^\nu} - \frac{1}{2} \frac{\partial g_{\mu\nu}}{\partial x^\lambda} \right) v^\mu v^\nu\right].
\end{equation}
% \textcolor{red}{can we absorb the minus sign into the first bracket?}
We can see that this equation has a $U^0$ term with the Newtonian potential, which is considered in \textsc{Hrna}. 

We use test cases for comparing both these techniques by using different setups of binary stars around a $10^6\;\mathrm{M}_\odot$ and $10^9\;\mathrm{M}_\odot$ black hole.

\subsection{Coordinate transformation: harmonic and Schwarzschild}
The harmonic coordinates can be converted to Schwarzschild coordinates using \citep{Sharpe2026}
\begin{align}
    \textbf{x}_\mathrm{sch} &= \left(1+\frac{R_\mathrm{sch}}{2 r_\mathrm{H}} \right)\mathbf{x}_\mathrm{H}\;,\\
    \textbf{v}_\mathrm{sch} &= \left(1+\frac{R_\mathrm{sch}}{2 r_\mathrm{H}} \right)\textbf{v}_\mathrm{H} -\frac{R_\mathrm{sch}}{2 r_\mathrm{H}^3}(\textbf{x}_\mathrm{H}\cdot\textbf{v}_\mathrm{H})\textbf{x}_\mathrm{H}\;,
\end{align}
where $R_\mathrm{sch}$ is the Schwarzschild radius and $r_\mathrm{H} \equiv |{x}_\mathrm{H}|$. Similarly, we can transform from Schwarzschild to harmonic using
\begin{align}
     \textbf{x}_\mathrm{H} &= \left(1-\frac{R_\mathrm{sch}}{2 r_\mathrm{sch}} \right)\mathbf{x}_\mathrm{sch}\;,\\
     \textbf{v}_\mathrm{H} &= \left(1-\frac{R_\mathrm{sch}}{2 r_\mathrm{sch}} \right)\textbf{v}_\mathrm{sch} +\frac{R_\mathrm{sch}}{2 r_\mathrm{sch}^3}(\textbf{x}_\mathrm{sch}\cdot\textbf{v}_\mathrm{sch})\textbf{x}_\mathrm{sch}\;,
\end{align}
where $r_\mathrm{sch} \equiv |{x}_\mathrm{sch}|$. This is derived from the following relation for radial component
\begin{align}
    r_\mathrm{sch} = r_\mathrm{H} + \frac{R_\mathrm{sch}}{2}\;.
\end{align}
Using these we map from \textsc{Multistar} to \textsc{Phantom-Geo} coordinates and vice-versa. The time coordinate stays the same. We assume that the Newtonian binary system is also in the Schwarzschild coordinate system which needs further exploration.

%eg - do we need this here? I think w should explain the statistics later. The results of individual cases should be presented next.

\section{Results}
\label{sec:results}

\subsection{Test 1: Binary around a $10^6\;\mathrm{M}_\odot$ SMBH}

Figure~\ref{fig:figure1} shows the orbital evolution of a binary placed at a distance of $50\; r_\mathrm{t}$
around a black hole of $10^6\;\mathrm{M}_\odot$. We use the same model as Figure $3$ of \citet{Manzaneda2024}, with initial conditions obtained from their \textsc{Hrna} run.  The binary has a semi-major axis of $a=0.01\;\mathrm{au}$ and total binary star mass ($m_b$) of $1\;\mathrm{M}_\odot$. This model uses $\beta \equiv r_\mathrm{t}/r_\mathrm{p}=5 \sim 20 \;r_\mathrm{g}$ and orbital phase of $\varphi = 1.41$ radian. We evolve this model for $1\;\mathrm{yr}$ in all codes. The coordinates obtained from \textsc{Phantom-Geo} and \textsc{Hrna} are converted to harmonic coordinates. The initial conditions set in the \textsc{PN} codes correspond to the initial harmonic coordinate values from the \textsc{Phantom-Geo} run.

\begin{figure*}
    \centering
    \includegraphics[width=\textwidth,height=\textheight, keepaspectratio]{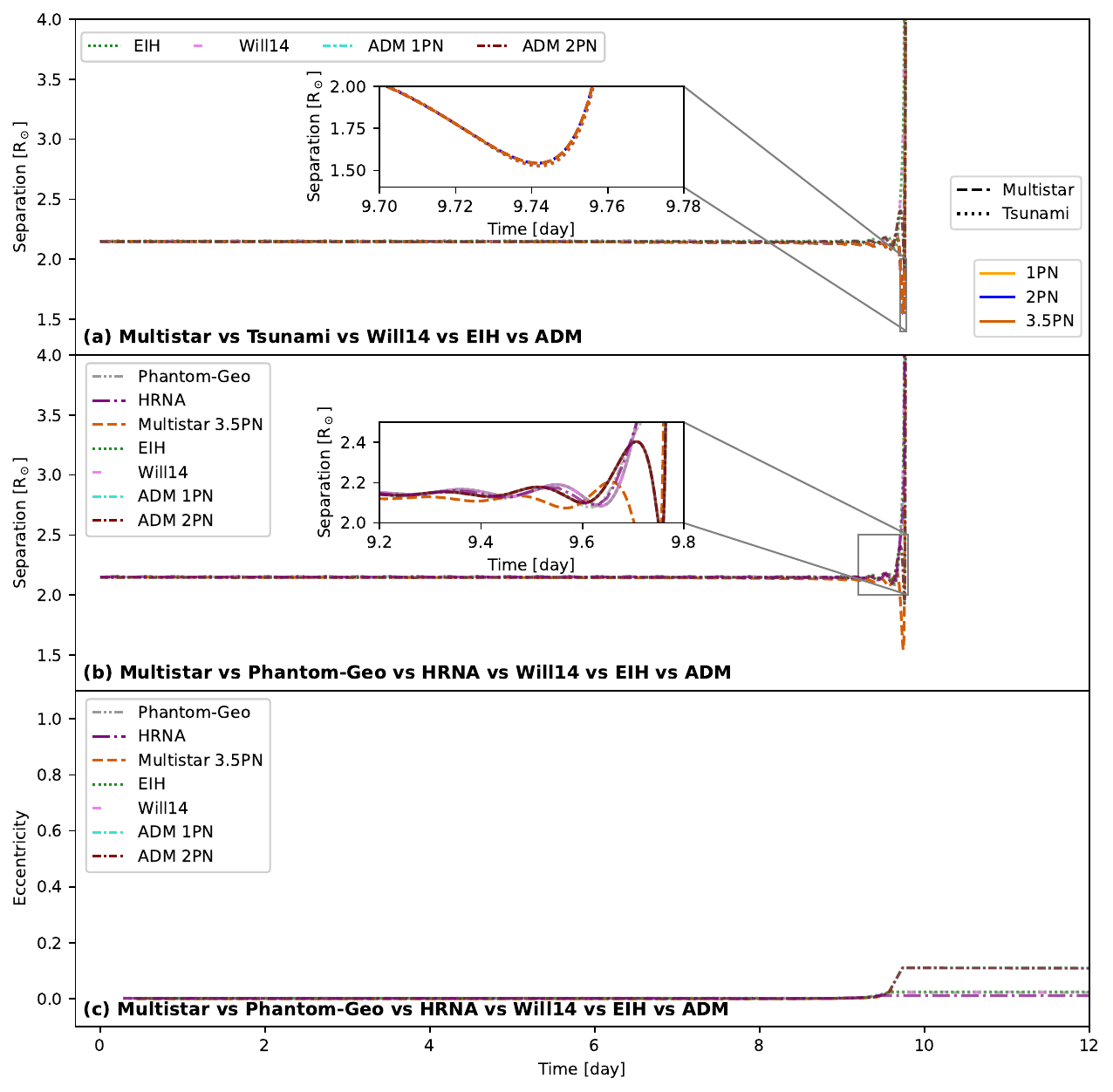}
    \caption{(a) Separation vs time for \textsc{Multistar}, \textsc{Tsunami}, \textsc{Will14}, \textsc{EIH} and \textsc{ADM} methods for model shown in Figure~\ref{fig:figure1}. The minimum separation in the pair-wise methods is $\sim 2 \;\mathrm{R}_\odot$. \textsc{Multistar}, \textsc{Tsunami} lines are over-plotted as they agree with each other. Similarly, \textsc{Will14} and \textsc{EIH} lie on top of each other. (b) Separation vs time for \textsc{Phantom-Geo} and \textsc{Hrna} compared with \textsc{PN} approaches. No methods show a change in separation similar to the pair-wise \textsc{PN} codes. (c) Eccentricity vs time shows that all models stay circular up to the pericentre. }
    \label{fig:figure1_panels}
\end{figure*}

Panel (a) shows the \textsc{Phantom-Geo} orbits of the two stars. Star $1$ gets bound to the SMBH while star $2$ is unbound. Panel (b) shows the comparison between \textsc{Phantom-Geo} and \textsc{Hrna}. Models agree with each other with $\sim1\,\%$ difference.  Panel (c) shows the comparison of \textsc{Phantom-Geo} with \textsc{EIH} and \textsc{Will14}. \textsc{EIH} and \textsc{Will14} amplitude match with each other with a difference of $\sim 0.02\,\%$, while about $8\,\%$ larger compared to \textsc{Phantom-Geo}. Phase of precession are not an exact match either. Panel (d) shows the comparison between the \textsc{Phantom-Geo} and \textsc{ADM} codes. $1\mathrm{PN}$ and $2\mathrm{PN}$ \textsc{ADM} both result in star 1 getting bound, and star 2 escaping. The amplitudes between the two formulations match within $0.5\,\%$, but the precession rates are different. $2\mathrm{PN}$ \textsc{ADM} is a better match to the \textsc{Phantom-Geo} code's precession rate, pointing towards convergence. Moreover, $1\mathrm{PN}$ \textsc{ADM} does not match with \textsc{EIH}, which might be due to different gauges.  2PN is closer to Geodesic than 1PN, suggesting that the results would converge at higher order. Panel (e) shows \textsc{Phantom-Geo} vs \textsc{Multistar}, where the latter is plotted up to $3.5\mathrm{PN}$. \textsc{Multistar}. $1\mathrm{PN}$ precession is slightly different from the \textsc{Phantom-Geo}, but the orientation matches when we add higher order terms, though the precession amplitudes are different by $\sim 20\,\%$. Also, opposite stars get bound and unbound in \textsc{Multistar} compared with \textsc{Phantom-Geo}. Panel (f) shows \textsc{Phantom-Geo} compared with the \textsc{Tsunami}. The two pair-wise codes match with each other with a difference of about $1\,\%$ in the amplitude of precessing orbits. But as the phase of the stars at the beginning of the simulation and the fact that this system is inherently chaotic, we can not be certain about convergence across codes.

Figure~\ref{fig:figure1_panels} shows the separation and eccentricity as function of time for the same model. The eccentricity of the binary is calculated by determining the trough and peak ($r_\mathrm{a}$ and $r_\mathrm{p}$) over each cycle using \textsc{scipy.find\_peaks}, and \begin{equation}
    e = \frac{r_\mathrm{a}-r_\mathrm{p}}{r_\mathrm{a} + r_\mathrm{p}}\;,
\end{equation}
as the Keplerian orbit eccentricity formula is not applicable in the GR case.  Panel~(a) shows the separation as function of time for the post-Newtonian methods. \textsc{Multistar} and \textsc{Tsunami} match with each other for different $\mathrm{PN}$. This model would not be flagged as a collision outcome in the \textsc{Multistar} code because the lowest separation is greater than sum of the radii of the two stars. Panel~(b) shows the separation of \textsc{Phantom-Geo} and \textsc{Hrna}. Separation never decreases to the point where the two stars would touch each other, but the stars split apart as was shown in Figure~\ref{fig:figure1}. The eccentricity evolution (Panel~c) shows that models remain circular as they approach the pericentre.

\begin{figure*}
    \centering
      \includegraphics[width=0.9\textwidth,height=0.9\textheight,keepaspectratio]{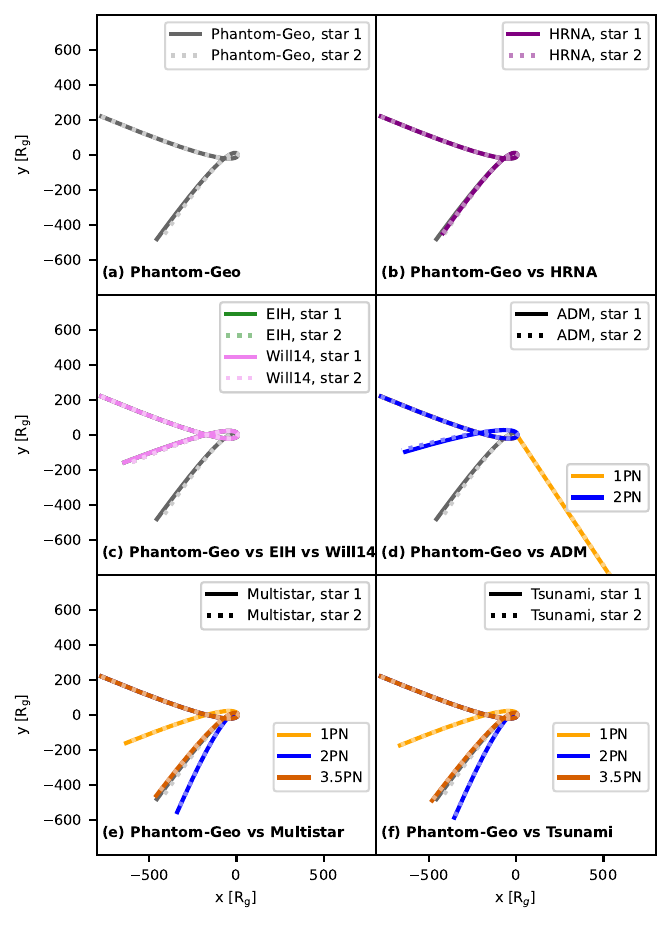}
    \caption{Comparison of models around $10^9\;\mathrm{M}_\odot$ black hole with $1\;\mathrm{M}_\odot$ stars, $0.1\;\mathrm{au}$ apart and $\beta=0.1$. The stars are split apart in \textsc{Phantom-Geo}, \textsc{Will14}, \textsc{EIH}, \textsc{Hrna}, \textsc{ADM} approaches but not in pair-wise $\mathrm{PN}$ codes. }
    \label{fig:figure_test1_all}
\end{figure*}

\begin{figure*}
    \centering
      \includegraphics[width=\textwidth,height=\textheight,keepaspectratio]{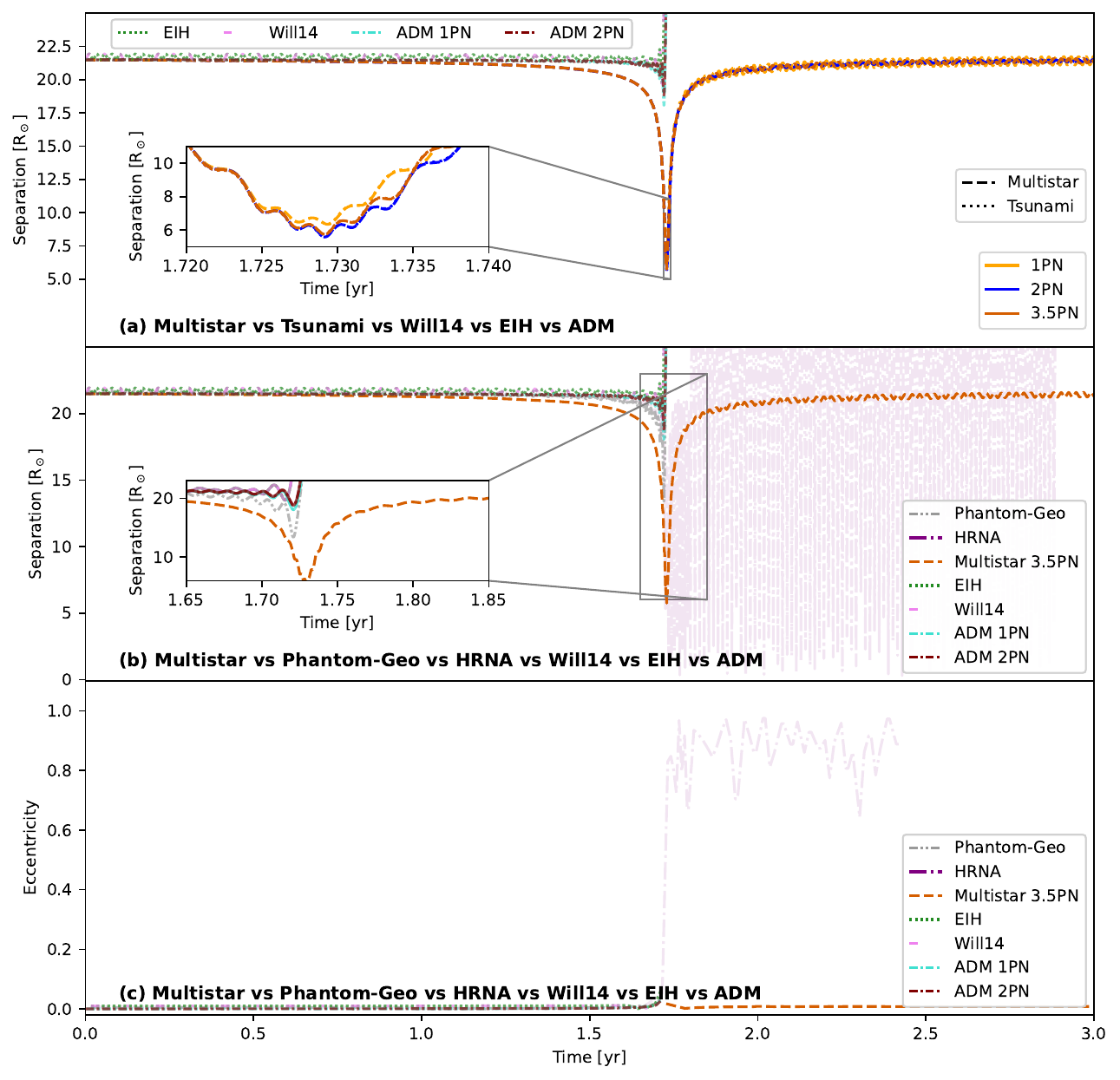}
    \caption{Separation and eccentricity of stars as function of time for the same model shown in Figure~\ref{fig:figure_test1_all}. Separation in the pair-wise \textsc{PN} codes, \textsc{Multistar} and \textsc{Tsunami} goes from $\sim 22\;\mathrm{R}_\odot$ to $\sim 5\;\mathrm{R}_\odot$, and the affect is reversible. The binary separation goes to $0\;\mathrm{R}_\odot$ in \textsc{Hrna} and the system becomes eccentric. We use a lighter shading for \textsc{Hrna} for this plot. }
    \label{fig:figure_sep_rad_test1}
\end{figure*}

\begin{figure*}
    \centering
      \includegraphics[width=0.9\textwidth,height=0.9\textheight,keepaspectratio]{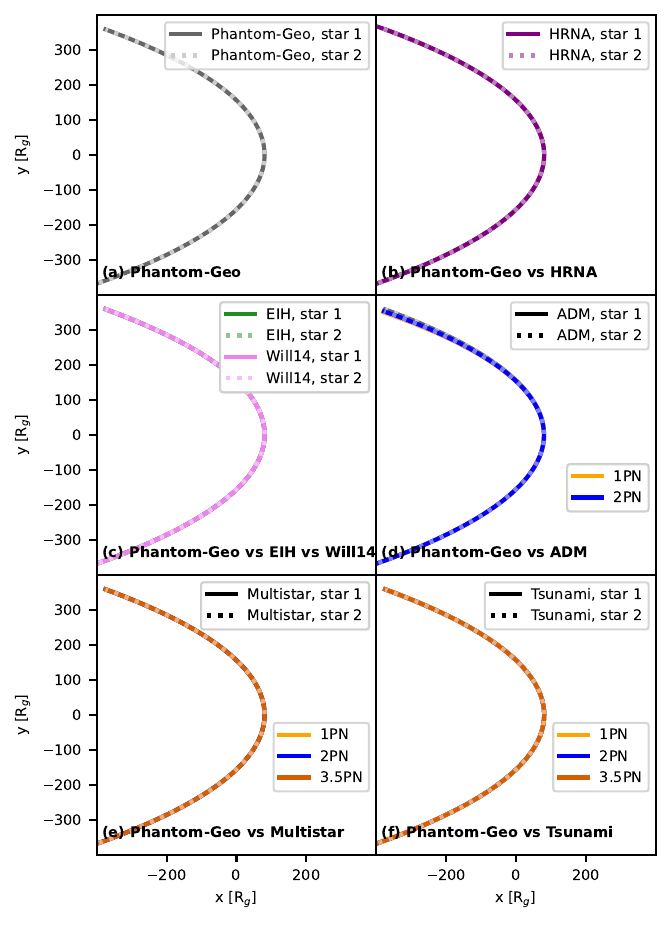}
    \caption{Orbits of stars for $\beta=0.01$ model evolved around a black hole of $10^9\;\mathrm{M}_\odot$. The stars follow similar orbits across different codes. }
    \label{fig:figure_beta01_1}
\end{figure*}

\begin{figure*}
    \centering
      \includegraphics[width=0.9\textwidth,height=0.9\textheight,keepaspectratio]{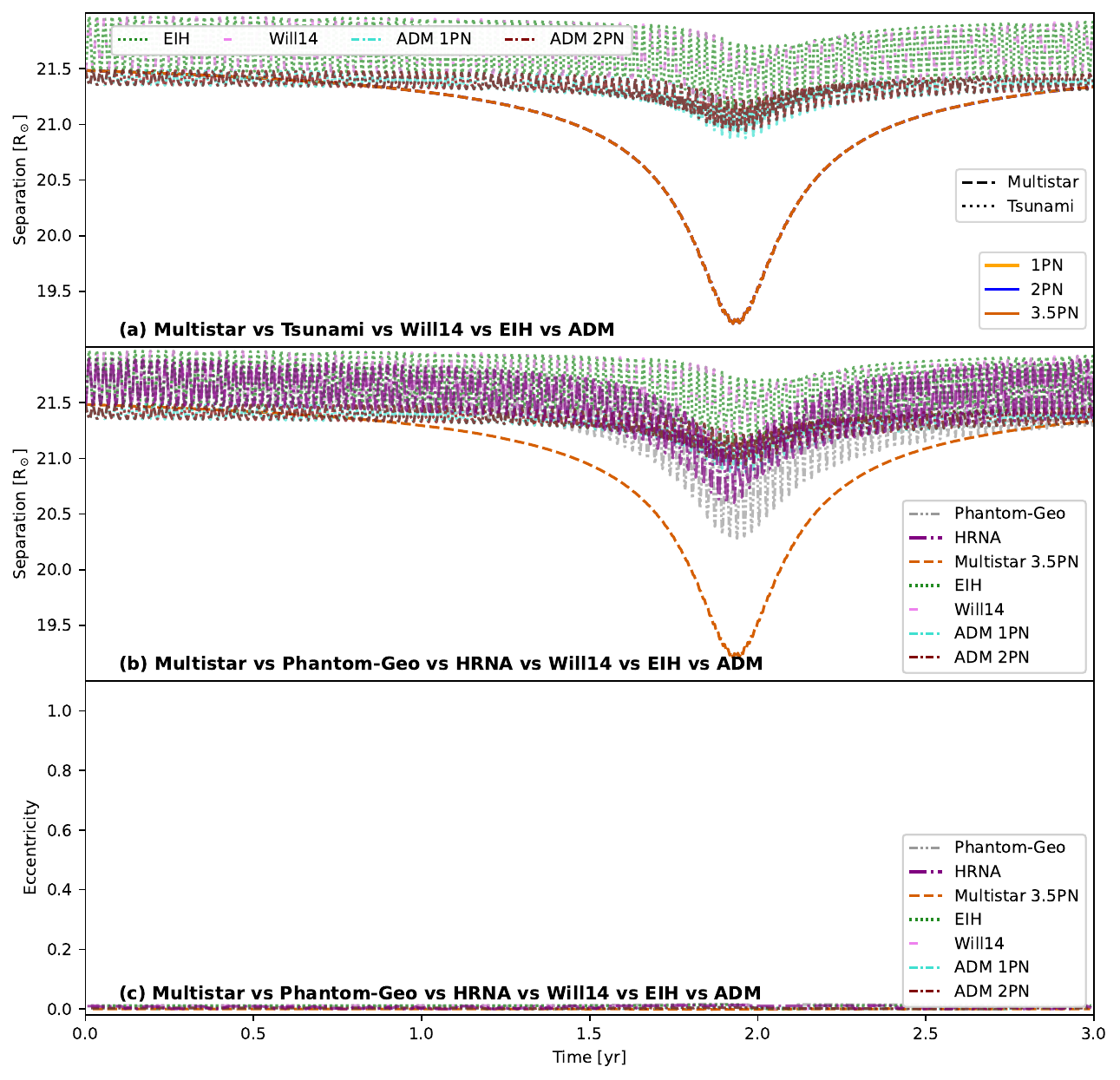}
    \caption{Separation and eccentricity for the model shown in Figure~\ref{fig:figure_beta01_1}. Pair-wise codes show the most decrease in separation amongst all codes. None of the codes result in eccentricity growth.}
    \label{fig:figure_beta01_2}
\end{figure*}

\subsection{Test 2: Binary around a $10^9\;\mathrm{M}_\odot$ SMBH}
\label{sec:test2}
\subsubsection{Strong GR case $(\beta=1)$}

We next consider a binary system of two $1\;\mathrm{M}_\odot$ stars orbiting a $10^9\;\mathrm{M}_\odot$ black hole, on an initially zero energy orbit (\ref{app:binary_setup}) in \textsc{Multistar} with $\beta=1$. The trajectory is started at $100\;r_\mathrm{t}$. We consider a semi-major axis of $0.1\;\mathrm{au}$, eccentricity, inclination, argument of periastron, and longitude of ascending node all set to \ang{0}, while mean anomaly of \ang{138.82}. This is a strong encounter as the pericentre is $\sim8 \;r_\mathrm{g}$. 

Fig~\ref{fig:figure_test1_all} shows the orbital evolution of the stars over $3$ years. In \textsc{Phantom-Geo}, the models split post-pericentre (first row, left panel). Compared with \textsc{Phantom-Geo}, the orbit precesses several degrees more in \textsc{EIH} and \textsc{Will14} formulations. In the \textsc{ADM} approach the stars evolve differently for $1\mathrm{PN}$ and $2\mathrm{PN}$ physics. \textsc{Multistar} and \textsc{Tsunami} orbital evolutions are similar to each other. The orbits match better with the \textsc{Phantom-Geo} as the order of $\mathrm{PN}$ is increased, with $3.5\mathrm{PN}$ being closer to the \textsc{Phantom-Geo} evolution, but the stars never split in the pair-wise codes but do in other schemes.

Figure~\ref{fig:figure_sep_rad_test1} panel (a) shows the separation between the binary stars in \textsc{Multistar} and \textsc{Tsunami}. We see a match between the two codes. \textsc{EIH} and \textsc{ADM} both result in binary stars splitting around the pericentre unlike pair-wise codes. The separation between the binary stars decreases by a factor of $4$ in the pair-wise codes whereas the other $\mathrm{PN}$ methods only cause a small decrease. Middle panel shows the separation evolution in \textsc{Phantom-Geo} and \textsc{Hrna}. Last panel shows the eccentricity evolution over three years for all codes. Except for \textsc{Hrna}, all codes only show a slight increase in eccentricity $< 0.05$, but that change is reversible in \textsc{Multistar}. In \textsc{Hrna}, the separation between the binary gets close to $0\;\mathrm{R}_\odot$ and the system becomes eccentric. This test shows that pair-wise \textsc{PN} codes result in reversible separation decrease, which is not present in other \textsc{PN} methods. Moreover, \textsc{Phantom-Geo} and \textsc{Hrna} do not match with each other for this test case.

%\subsubsection{Test 2}
\subsubsection{Weak GR case $(\beta \ll 1)$}We now consider similar binary setup as before, but with $\beta=0.01$ ($\sim 800 \;r_\mathrm{g}$) and mean anomaly of \ang{133.71}, so that the binary experiences very weak force from the SMBH. Figure~\ref{fig:figure_beta01_1} shows the trajectories of binary stars. Panel (a) shows \textsc{Phantom-Geo} where both stars stay bound post-disruption. The same behaviour is observed in all the codes as well. Figure~\ref{fig:figure_beta01_2} shows the separation and eccentricity evolution for different codes. The decrease in separation is the largest ($\sim 12\%$) in \textsc{Multistar} and is also present in \textsc{Tsunami}. This behaviour is reversible, with only a slight increase in eccentricity close to the pericentre ($< 0.02$). In \textsc{Phantom-Geo} separation decreases by $\sim4\%$ around pericentre, with $\sim3\%$ change in \textsc{Hrna}, and $\sim2\%$ change in other \textsc{PN} schemes. None of the models become eccentric. Overall, a low $\beta$ results in similar behaviour across different schemes, but pair-wise \textsc{PN} results in the largest decrease in separation.

\section{Discussion}
\label{sec:discussion}
\subsection{Simulating statistics of outcomes around $10^6 \;\mathrm{M}_\odot$ black hole}

The inherently chaotic nature of the three body problem \citep{Mardling2008,Zwart2022}, the different methods and coordinate systems and numerical round-off error limit our ability to simulate the exact same system, hence a statistical approach is required to adequately estimate the trends and outcomes of different methods.
%\EG{[Tried to improve the grammar and flow of the first paragraph here.]}

%We perform statistics because small differences in the inner or other orbit physics due to the different approaches, or even numerical round-off in the initial conditions, could result in very different results.  And different approaches, as we have discovered, make it even difficult to ensure that all models are started with really exactly the same initial conditions.  The variations then are because the three-body system is chaotic \citep{Mardling2008,Zwart2022}. 

To understand if the different approaches result in same statistical outcomes, we perform a study using both \textsc{Phantom-Geo} and \textsc{Multistar}, similar to \citet{Manzaneda2024}. We consider $200$ different initial conditions, uniformly sampled in $\beta \in [0.1, 10]$ with phase of the binary $\phi \in [0, \pi]$ and only considering retrograde orbits. We generate a grid of $400$ phase values, which implies $80\mathord,000$ models for each code.  We consider $0.5 + 0.5\;\mathrm{M}_\odot$ binary system around a $10^6\;\mathrm{M}_\odot$ black hole, with eccentricity of zero and semi-major axis of $0.01\;\mathrm{au}$. We consider $80\mathord,000$ zero energy orbits of the binary around the black hole (see \ref{app:binary_setup}) and plot the fraction of surviving models. We determine survival of binary in all the codes by calculating the energy of the binary in its frame by using
\begin{equation}
    \varepsilon =  \frac{1}{2}\frac{m_i m_j}{m_i + m_j} v_{ij}^2 - \frac{G  m_i m_j}{r_{ij}}\;,
\end{equation}
where $\varepsilon<0$ implies survival. All models start at $50\;r_\mathrm{t}$ and terminate at $150\;r_\mathrm{t}$. We also compare with \textsc{Tsunami} and \textsc{Hrna}. 
with semi-major axis of $a$, and mass of binary of $m_\mathrm{b}$. 

\begin{figure}
    \centering
    \includegraphics[width=\columnwidth]{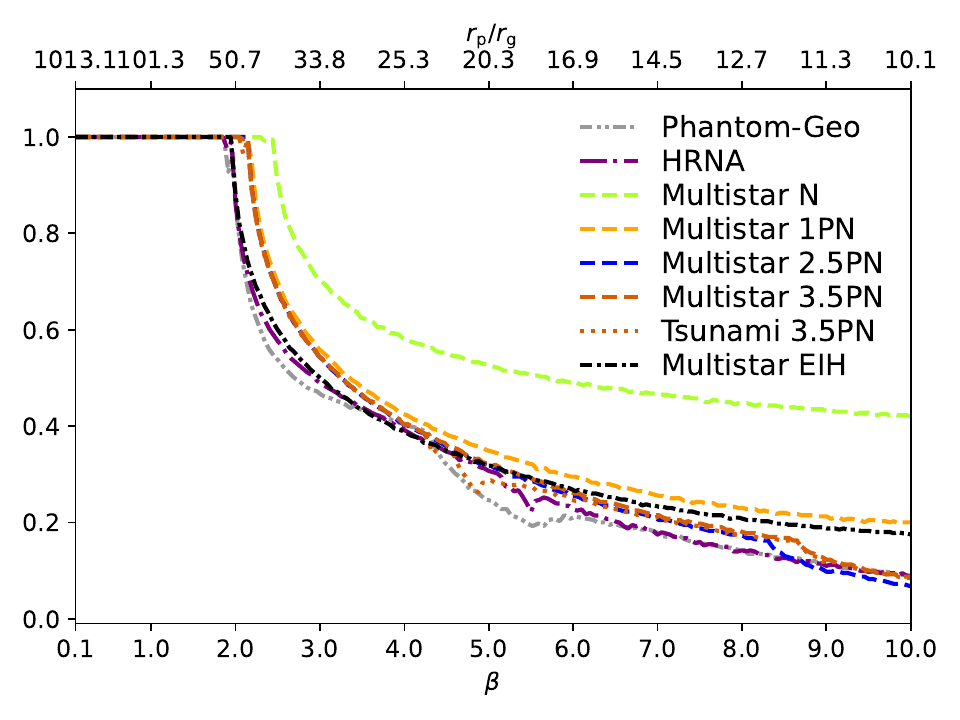}
    \caption{Fraction of binaries that survive the encounter with a $10^6\;\mathrm{M}_\odot$ black hole. We run the models \textsc{Phantom-Geo}, \textsc{Hrna}, \textsc{Multistar} and \textsc{Tsunami}. Higher order $\mathrm{PN}$ pair-wise codes match with the metric-with-perturbation approaches. }
    \label{fig:systems_rate}
\end{figure}

We consider the pair-wise $\mathrm{PN}$, \textsc{EIH} (\ref{app:eih_multi}) and metric-with-perturbation methods for comparison. Figure~\ref{fig:systems_rate} shows the fraction for \textsc{Phantom-Geo}, \textsc{Hrna}, \textsc{Multistar} pair-wise, \textsc{Multistar} \textsc{EIH} and \textsc{Tsunami} codes. Newtonian gravity results in models surviving up to $\beta \sim 3$, and the fraction of survivals is always larger compared with GR schemes. For \textsc{Multistar} $1\mathrm{PN}$, $2.5\mathrm{PN}$ and $3.5$ all models survive up to $\beta \sim 2.07$, while for \textsc{EIH} the boundary is $\sim 1.89$. \textsc{EIH} behaviour is closer to the \textsc{Phantom-Geo} and \textsc{Hrna} for which all models survive up to the same $\beta$. We note that around $\beta=5$, there is a decrease in survival fraction in \textsc{Phantom-Geo}. A similar behaviour is present in \textsc{Hrna} and \textsc{Tsunami}, but this needs further exploration. We postulate that the decrease may be a result of resonances or other high-order effects in the system. However, detailed study of this phenomena is beyond the scope of this paper. %combined with small statistical sampling 
 Overall, \textsc{Tsunami} and \textsc{Multistar} pair-wise methods behave similarly, and so do \textsc{Phantom-Geo} and \textsc{Hrna}. Close to $\beta \sim 10$, only a small fraction ($0.1$) of binaries survive in the pair-wise \textsc{PN} and metric-with-perturbation codes. \textsc{Multistar} EIH has a fraction of $\sim 0.18$ for $\beta=10$.

\subsubsection{Classification of binary events around $10^9\;\mathrm{M}_\odot$ black hole in \textsc{Multistar}}
We next explore the zero-energy orbits in \textsc{Multistar}. We consider a binary of $0.5\;\mathrm{M}_\odot$ stars, with $0.01\;\mathrm{au}$. Again we only consider retrograde orbits, but this time we classify the outcomes into collision (binary separation $<$ sum of radii) and binary capture around the black hole. We find that all $1\mathord,000$ test models collide with each other due to the decrease in separation. This might be an artefact of the implementation.

We next increase the binary semi-major axis to $0.1\;\mathrm{au}$ in \textsc{Multistar} and find that the systems survive and get captured for low $\beta$ values, though there is slight decrease in separation close to the pericentre. This is consistent with what we observed in tests shown in Figure~\ref{fig:figure_test1_all} to Figure~\ref{fig:figure_beta01_2}. Hence, if one wishes to perform three-body problem around a black hole using a \textsc{PN} scheme, we suggest to avoid using the pair-wise \textsc{PN} method.

\subsection{Which method to trust?}

We used different formulations for solving the three-body problem in general relativity around two different black hole masses.  The tests of $0.5\;\mathrm{M}_\odot$ binary stars placed $0.01\;\mathrm{au}$ apart around the $10^6\;\mathrm{M}_\odot$ black hole shows that the different approaches are consistent with each other for $\beta \sim 5$, but diverge afterwards. When we place $1\;\mathrm{M}_\odot$ stars $0.1\;\mathrm{au}$ apart around a $10^9\;\mathrm{M}_\odot$ black hole, we find that different approaches are consistent with each othe  when the binary is placed at a $\beta \sim 0.1$, while at $\beta\sim 1.0$, the binary is disrupted in the \textsc{Phantom-Geo}, \textsc{EIH}, \textsc{Will14} and \textsc{ADM} approaches, but not in the pair-wise scheme. We also find that the pair-wise scheme results in decrease in separation close to the pericentre in all test cases. 

We conclude that the pair-wise $\mathrm{PN}$ scheme is the least reliable for this problem, due to the decrease in separation that takes place even for a low impact parameter case, with the system never becoming eccentric. This is a feature of the approach, and not a numerical artefact as the behaviour is present in both \textsc{Multistar} and \textsc{Tsunami}.

 In \textsc{Phantom-Geo} we ignored the $U^0$ term in the Equation (\ref{eq:mom_eq}) while \textsc{Hrna} has it in Equation (\ref{eq:hmra_vel}). We tested all out models by reintroducing all the terms we ignored in Equation (\ref{eq:mom_eq}). We found that the angular momentum conservation became worse ($\sim10^{-4}$) and hence, we ignored those results in this paper. Though the outcomes of our reintroduced runs became closer to \textsc{Multistar} for the $10^9\;\mathrm{M}_\odot$ black hole, Test $1$.  \textsc{Hrna} also has poor angular momentum and energy conservation when we run these models which are close to the speed of light and close to the black hole.  If we do not use any correction terms in Equation (\ref{eq:mom_eq}), angular momentum and energy were conserved up to the round-off error. We also tested a different time-stepping algorithm (Heun's method) and found the same outcome in \textsc{Phantom-Geo}. 

We note that the \textsc{ADM} approach uses a different gauge, and we did not perform any coordinate transformation in this work for the \textsc{ADM} case. Also, the pair-wise $\mathrm{PN}$ method might not hold for the harmonic gauge condition, as we add the pair-wise terms for three bodies. 

Overall, both post-Newtonian, and the metric-with-\\perturbation schemes are consistent with each other when we consider a wide binary, and a system that is moving at velocity $< 0.35\;c$. Using pair-wise $\mathrm{PN}$ is not a reliable method for solving the three-body problem due to the artificial decrease in separation close to the pericentre even for low $\beta$ values. Furthermore, pair-wise codes do not consider cross terms which have been shown to be important. As all methods assume a weak field scenario, care needs to be taken when interpreting the results close to the black hole. We propose that the best method for solving $N$-body problem is to use would be either \textsc{EIH} or \textsc{ADM} approaches, followed by metric-with-perturbation scheme.

As was shown by \citet{Manzaneda2024}, as GR affects the velocities of the hypervelocity stars, incorporating GR by these approximations is valuable for such studies. Similar consideration is required for understanding double tidal disruption event rates to obtain accurate statistics. Furthermore, when using $N$-body codes which use pair-wise $\mathrm{PN}$ implementation, consideration is needed when stars move close to the speed of light. Our results are similar to \citet{Zwart2022} who found that for $v/c \sim 0.4$, the \textsc{PN} methods do not capture the full physics, and pair-wise \textsc{PN} for $N > 3$ behaves differently than when cross-terms are taken into account. 

\subsection{Future work}

Future work would involve solving the Einstein equations for the three-body system in \textsc{Phantom-Geo}, for example solving in linearised gravity with retarded potentials to account for the relativistic effects on the binary as it gets closer to the black hole. This might resolve the decrease in separation noted in different formulations.

Moreover, obtaining the $2\mathrm{PN}$ 3-body equations in harmonic coordinates would be an interesting problem to solve. We are currently working on obtaining collision statistics of binary stars around black holes of different masses (Sharma et al. in prep).

\section{Conclusion}
\label{sec:conclusion}

We compared 7 different ways to solve the three-body problem around two black hole masses of $10^6\;\mathrm{M}_\odot$ and $10^9\;\mathrm{M}_\odot$. We found that:
\begin{enumerate}
    \item Pair-wise and \textsc{EIH} $\mathrm{PN}$ schemes and metric-with-perturbation codes match for interactions of binary stars with a $10^6\,\mathrm{M_\odot}$ black hole. 
    \item Statistically, higher order pair-wise $\mathrm{PN}$ codes match better with the metric-with-perturbation codes.
    \item  Pair-wise $\mathrm{PN}$ methods result in an artificial decrease in separation close to the pericentre, making this the least reliable method of all. This effect becomes more prominent around $10^9\;\mathrm{M}_\odot$ black holes as the binary separation decreases ($\leq 0.1\;\mathrm{au}$).
    \item We suggest the use of \textsc{EIH}, \textsc{Will14} and \textsc{ADM} as the most reliable approach for this problem, but consideration is required because these methods assume weak field interactions.
\end{enumerate}

\section*{Acknowledgements}
We thank Bernhard Mueller, Jordan Moncrieff and Charlie Sharpe for helpful discussions. DP also thanks Alessandro Lupi for useful discussions. We utilised \textsc{sympy} \citep{sympy}, \textsc{numpy} \citep{Numpy2020}, \textsc{matplotlib} \citep{Hunter2007}, and \textsc{scipy} \citep{Scipy2020} for analysis and plotting. EG acknowledges support from the ARC Discovery Early Career Research Award (DECRA) DE260101802 and the ARC Discovery Program DP240103174 (PI: Heger). MS acknowledges Monash University for the Monash Graduate Scholarship and Monash International Tuition Scholarship.

%%%%%%%%%%%%%%%%%%%%%%%%%%%%%%%%%%%%%%%%%%%%%%%%%%
\section*{Data Availability}
Our model initial conditions, and output files will be available on publication at \url{https://dx.doi.org/10.5281/zenodo.19703205}.

\appendix

\section{Two-body $3\mathrm{PN}$ and $3.5\mathrm{PN}$ terms}
\label{app:post_new_terms}
The $3\mathrm{PN}$ term (\citealt{Blanchet2001}, \citealt{Blanchet2024}, Equations (357), (362a), and (362b)) is 
\begin{align*}
    \mathbf{F}_{ij,3\mathrm{PN}} &= -\frac{G m_i m_j}{r_{ij}^2 c^6} \left\{ 
    \left[ 
    \left( -\frac{35}{16} + \eta \left(1-\eta\right) \frac{175}{16}\right)\eta \dot{r}_{ij}^6 \right. \right. \nonumber \\
    &\quad \left. \left. 
    +\left(\frac{15}{2} + \eta \left(- \frac{135}{4} + \frac{255 \eta}{8}\right)\right)\eta \dot{r}_{ij}^4 v^2_{ij}
    \right. \right. \nonumber \\
    &\quad \left. \left. 
    +\left(-\frac{15}{2} + \eta \left(\frac{237}{8} - \frac{45\eta}{2}\right) \right)\eta \dot{r}^2_{ij}  v^4_{ij} 
    \right. \right. \nonumber \\
    &\quad \left. \left. 
    +\left( \frac{11}{4} + \eta\left(-\frac{49}{4} + 13 \eta\right) \right)\eta v^6_{ij}
    \right. \right. \nonumber \\
    &\quad \left. \left. 
    +\frac{G m_{ij}}{r_{ij}} \left(79 - \eta \left(\frac{69}{2} + 30 \eta\right)\right)\eta \dot{r}^4_{ij} 
    \right. \right. \nonumber \\
    &\quad \left. \left. 
    +\frac{G m_{ij}}{r} \left(-121 + \eta \left(16 + 20 \eta\right)\right)\eta \dot{r}^2_{ij} v^2_{ij} 
    \right. \right. \nonumber \\
    &\quad \left. \left. 
    +\frac{G m_{ij}}{r_{ij}} \left( \frac{75}{4} + \eta\left(8-10\eta\right)\right)\eta v^4_{ij} 
    \right. \right. \nonumber \\
    &\quad \left. \left. 
    +\frac{G^2 m_{ij}^2}{r^2_{ij}} \left(1 + \eta\left(\frac{22717}{168} + \frac{615\pi^2}{64}\right)
    \right. \right. \right. \nonumber \\
    &\quad \left. \left. \left.
    + \eta^2\left(\frac{11}{8} - 7\eta\right)\right)\dot{r}^2_{ij}
    \right. \right. \nonumber \\
    &\quad \left. \left. 
    +\frac{G^2 m_{ij}^2}{r^2_{ij}} \left( \eta^2 - \frac{20827}{840} - \frac{123\pi^2}{64}\right) \eta v^2_{ij} 
    \right. \right. \nonumber \\
    &\quad \left. \left. 
    +\frac{G^3 m_{ij}^3}{r^3_{ij}}\left(-16 -\eta\left(\frac{1399}{12} - \frac{41\pi^2}{16} + \frac{71\eta}{2}\right) \right)
    \right] \mathbf{n}_{ij} \right. \nonumber \\
    &\quad \left. 
    +\left[
    \eta \left( -\frac{45}{8} + \left(15 + \frac{15\eta}{4}\right)\right) \dot{r}^5_{ij} \right. \right. \nonumber \\
    &\quad \left. \left. 
    + \eta \left(12 + \eta\left(-\frac{111}{4} - 12 \eta \right) \right) \dot{r}^3_{ij}
 v^2_{ij}  \right. \right. \nonumber \\
&\quad \left. \left. 
 + \eta \left(-\frac{65}{8} + \eta \left(19 + 6 \eta \right)\right)  \dot{r}_{ij} v^4_{ij} \right. \right. \nonumber \\
 &\quad \left. \left. 
 + \frac{G m_{ij}}{r_{ij}}\eta \left( \frac{329}{6} + \eta\left( \frac{59}{2} + 18 \eta\right) \right) \dot{r}^3_{ij} \right. \right. \nonumber \\
 &\quad \left. \left.
 -\frac{G m_{ij}}{r_{ij}} \eta \left( 15 + \eta \left( 27 + 10 \eta\right) \right) v^2_{ij} \dot{r}_{ij} \right. \right. \nonumber \\
&\quad \left. \left. 
+ \frac{G^2 m_{ij}^2}{r^2_{ij}}\left( -4 - \eta \left( \frac{5849}{840} + \frac{123\pi^2}{32} \right) 
\right. \right. \right. \\
&\quad \left. \left. \left.
+ \eta^2 \left(25 + 8\eta \right) \right) \dot{r}_{ij}
 \right] \mathbf{v}_{ij} 
    \right\}
\end{align*}
% \textcolor{red}{can we compress these formulae a bit - fewer lines?  Some journals offer typeset of long formulae as single-column (see Kidder 1995).  Can we do this here?}
% \textcolor{purple}{I tried using cuted package, and use strip, but it makes equations forced on same page, so they get cut to fit the same page, even though they should move to next page to fit }
The $3.5\mathrm{PN}$ term (\citealt{Iyer1993}, \citealt{Blanchet2024}, Equations (357), (358e), and (359e)) is 
\begin{align}
    \mathbf{F}_{ij,3.5\mathrm{PN}} &= -\frac{G m_i m_j}{r^2_{ij} c^7} \left\{ 
    \frac{G m_{ij} \eta \dot{r}_{ij}}{r_{ij}}
    \left[ \frac{G^2 m_{ij}^2}{r^2_{ij}} \left( \frac{3956}{35} + \frac{184\eta}{5}\right) \right. \right. \nonumber \\
    &\quad \left. \left.
    + \frac{G m_{ij} v^2_{ij}}{r_{ij}} \left( \frac{692}{35} - \frac{724\eta}{15}\right)
    + v^4_{ij} \left(\frac{366}{35} + 12 \eta \right) \right. \right. \nonumber\\
    &\quad \left. \left. 
    + \frac{G m_{ij} \dot{r}^2_{ij}}{r}\left(\frac{294}{5} + \frac{376\eta}{5} \right) \right. \right. \nonumber \\
  &\quad \left. \left. 
  - v^2_{ij} \dot{r}_{ij}^2\left(114 + 12 \eta \right) + 112 \dot{r}^4_{ij}
    \right]\textbf{n}_{ij} \right. \nonumber\\
    & \quad \left. 
   + \frac{G m_{ij} \eta}{r_{ij}} \left[ -\frac{G^2 m_{ij}^2}{r^2_{ij}} \left( \frac{1060}{21} + \frac{104\eta}{5}\right) \right. \right. \nonumber\\
   &\quad \left. \left. 
   + \frac{G m_{ij}\eta^2}{r_{ij}}\left(\frac{164}{21} + \frac{148\eta}{5} \right) 
   + v^4_{ij} \left(-\frac{626}{35} - \frac{12\eta}{5} \right) \right. \right. \nonumber \\
   &\quad \left. \left.
    + \frac{G m_{ij}\dot{r}_{ij}^2}{r_{ij}}\left(-\frac{82}{3} - \frac{848\eta}{15} \right) \right. \right. \nonumber\\
    &\quad \left. \left. 
     + v^2_{ij} \dot{r}_{ij}^2 \left(\frac{678}{5} + \frac{12\eta}{5} \right) -120 \dot{r}^4_{ij}
    \right]\textbf{v}_{ij}
    \right\}\;.
\end{align}

\section{ADM three-body $2\mathrm{PN}$ Hamiltonian term}
\label{app:ADM2pn}
The $2\mathrm{PN}$ Hamiltonian is given by (\citet{Galaviz2011}, their Equation~A2)
\begin{align}
\label{eq:adm2pn_eq}
    H_2 &= \frac{1}{16} \left(\frac{\mathbf{p}_i^2}{m_i^2}\right)^3 + \frac{1}{16}\sum_i \sum_{j\neq i} \frac{G m_i^{-1}m_j^{-1}}{r_{ij}}\left[10 \left( \frac{m_j}{m_i} \mathbf{p}_i^{\,2} \right)^2\right.\nonumber\\
   &\quad \left. - 11 \mathbf{p}_i^{\,2} \mathbf{p}_j^{\,2}
   - 2 (\mathbf{p}_i \!\cdot\! \mathbf{p}_j)^2
   + 10 \mathbf{p}_i^{\,2} (\hat{ \mathbf{n}}_{ij}\!\cdot\!\mathbf{p}_j)^2\right.\nonumber\\
   &\quad \left.- 12 (\mathbf{p}_i \!\cdot\! \mathbf{p}_j) (\hat{ \mathbf{n}}_{ij}\!\cdot\!\mathbf{p}_i)(\hat{\mathbf{n}}_{ij}\!\cdot\!\mathbf{p}_j) \right. \nonumber\\
  &\quad \left. - 3 (\hat{\mathbf{ n}}_{ij}\!\cdot\!\mathbf{p}_i)^2 (\hat{\mathbf{n}}_{ij}\!\cdot\!\mathbf{p}_j)^2\right]+ \frac{1}{8} \sum_i \sum_{j\neq i} \sum_{k\neq i}
   \frac{G^2}{r_{ij} r_{ik}}\left[
   \frac{18 m_j m_k}{m_i}\mathbf{p}_i^2\right.\nonumber\\
   &\quad \left. + 14 \frac{m_i m_k}{m_j}\mathbf{p}_j^2 - 2 \frac{m_i m_k}{m_j}(\hat{\mathbf{n}}_{ij}\!\cdot\!\mathbf{p}_j)^2 - 50 m_k (\mathbf{p}_i \!\cdot\!\mathbf{p}_j) + 17 m_i (\mathbf{p}_i\!\cdot\!\mathbf{p}_k)\right. \nonumber\\
   &\quad \left.-14 m_k (\hat{\mathbf{n}}_{ij}\!\cdot\!\mathbf{p}_i)(\hat{\mathbf{n}}_{ij}\!\cdot\!\mathbf{p}_j) + 14 m_i (\hat{\mathbf{n}}_{ij}\!\cdot\!\mathbf{p}_i)(\hat{\mathbf{n}}_{ij}\!\cdot\!\mathbf{p}_k) \right. \nonumber\\
   &\quad \left. + m_i (\hat{\mathbf{n}}_{ij}\!\cdot\hat{\mathbf{n}}_{ik}) (\hat{\mathbf{n}}_{ij}\!\cdot\!\mathbf{p}_{j}) (\hat{\mathbf{n}}_{ij}\!\cdot\!\mathbf{p}_k)
   \right] \nonumber\\
   &\quad + \frac{1}{8}\sum_i \sum_{j\neq i} \sum_{k\neq i} \frac{G^2}{r_{ij}^2} \left[2 m_j (\hat{\mathbf{n}}_{ij}\!\cdot\!\mathbf{p}_i)(\hat{\mathbf{n}}_{ik}\!\cdot\!\mathbf{p}_k)
  \right. \nonumber \\
   &\quad \left. + 2 m_j (\hat{\mathbf{n}}_{ij}\!\cdot\!\mathbf{p}_j)(\hat{\mathbf{n}}_{ik}\!\cdot\!\mathbf{p}_k)+\frac{m_i m_j}{m_k}(5 (\hat{\mathbf{n}}_{ij}\!\cdot\!\hat{\mathbf{n}}_{ik})\mathbf{p}_k^2 \right. \nonumber \\
   &\quad \left. - (\hat{\mathbf{n}}_{ij}\!\cdot\!\hat{\mathbf{n}}_{ik})(\hat{\mathbf{n}}_{ij}\!\cdot\!\mathbf{p}_k)^2) -14 (\hat{\mathbf{n}}_{ij}\!\cdot\!\mathbf{p}_k)(\hat{\mathbf{n}}_{ik}\!\cdot\!\mathbf{p}_k)) \right]  \nonumber\\
   &\quad +\frac{1}{4}\sum_i \sum_{j\neq i} \frac{G^2 m_i}{r_{ij}^2} \left[\frac{m_j}{m_i} \mathbf{p}_i^2 + \frac{m_i}{m_j}\mathrm{p}_j^2 - 2(\mathbf{p}_i\!\cdot\!\mathbf{p}_j) \right] \nonumber\\
   &\quad +\frac{G^2}{2}\sum_{i}\sum_{j\ne i}\sum_{k \ne i, j} \frac{(n_{ij}^a + n_{ik}^a)(n_{ij}^b + n_{kj}^b)}{(r_{ij} + r_{jk} + r_{ki})^2} \left[8 m_j (p_{ia}p_{kb})\right.\nonumber\\&\quad -16 m_j (p_{ib}p_{ka}) + 3 m_k (p_{ia}p_{jb}) + 4 \frac{m_i m_j}{m_k}(p_{ka}p_{kb}) + \frac{m_j m_k}{m_i} (p_{ia}p_{ib})  \left.\right] \nonumber \\
   &\quad + \sum_i \sum_{j\ne i} \sum_{k \ne i,j}\frac{G^2 m_i m_j m_k}{(r_{ij}+r_{jk}+r_{ki})r_{ij}} \left[8 \frac{\mathbf{p}_i \!\cdot\!\mathbf{p}_k - (\hat{\mathbf{n}}_{ij}\!\cdot\!\mathbf{p}_i)(\hat{\mathbf{n}}_{ij}\!\cdot\!\mathbf{p}_k)}{m_i m_k} \right. \nonumber \\
   &\quad \left. - 3 \frac{\mathbf{p}_i\!\cdot\!\mathbf{p}_j - (\hat{\mathbf{n}}_{ij}\!\cdot\!\mathbf{p}_i) (\hat{\mathbf{n}}_{ij}\!\cdot\!\mathbf{p}_j)}{m_i m_j} - 4 \frac{\mathbf{p}_k^2 - (\hat{\mathbf{n}}_{ij}\!\cdot\!\mathbf{p}_k)^2}{m_k^2} - \frac{\mathbf{p}_i^2 - (\hat{\mathbf{n}}_{ij}\!\cdot\!\mathbf{p}_i)^2}{m_i^2}\right]
   \nonumber \\
   &\quad -\frac{G^3}{2}\sum_i\sum_{j\ne i}\left(\sum_{k\ne i,j} \frac{m_i^2 m_j m_k}{r_{ij}^2 r_{jk}} + \frac{1}{2} \sum_{k\ne j}\frac{m_i^2 m_j m_k}{r_{ij}^2 r_{jk}}\right) \nonumber \\
   &\quad - \frac{3G^3}{8}\sum_i \sum_{j \ne i} \left(\sum_{k \ne i} \frac{m_i^2 m_j m_k}{r_{ij}^2 r_{ik}} + \sum_{k\ne i,j} \frac{m_i^2 m_j m_k}{r_{ij}^2 r_{ik}} \right)\nonumber\\
   &\quad - \frac{3 G^3}{8} \sum_i \sum_{j \ne i} \sum_{k\ne i,j} \frac{m_i^2 m_j m_k}{\textcolor{red}{r_{ij}} r_{ik} r_{jk}} \nonumber\\
   &\quad - \frac{G^3}{64} \sum_i \sum_{j \ne i} \sum_{k \ne i,j} \frac{m_i^2 m_j m_k}{r_{ij} r_{ik}^3 r_{jk}} \left( 18 r_{ik}^2 - 60 r_{jk}^2 - 24 r_{ik} (r_{ij}+ r_{jk})\right. \nonumber \\
  &\quad  \left. +  60 \frac{r_{ik} r_{jk}^2}{r_{ij}} + 56 r_{ij} r_{jk} - 72 \frac{r_{jk}^3}{r_{ij}} + 35 \frac{r_{jk}^4}{r_{ij}^2} + 6 r_{ij}^2\right) \nonumber\\
  &\quad - \frac{G^3}{4} \sum_i \sum_{j\ne i} \frac{m_i^2 m_j^2}{r_{ij}^3}\;. 
\end{align}

\section{\textsc{Phantom-Geo} test case: binary system in isolation}
\label{app:geodesic_test}
\begin{figure}
    \centering
    \includegraphics[width=\columnwidth]{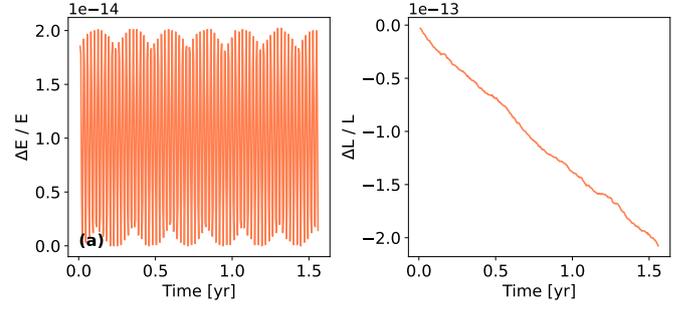}
    \caption{a) The fractional change in energy as a function of time b) the fractional change in angular momentum as a function of time. Both quantities agree within the round-off error.}
    \label{fig:binary_star_energy}
\end{figure}

We test that our binary stars would evolve normally in the Minkowski metric.  We consider a binary stars system with $1\;\mathrm{M}_\odot$ stars, a semi-major axis of $0.1\;\mathrm{au}$, and a zero eccentricity. The system stays circular over $100$ orbits. We use a numerical precision of up to $8$ decimal places. Figure~\ref{fig:binary_star_energy} shows the energy and angular momentum error for our model. 

\section{Binary system setup for statistics}
\label{app:binary_setup}
\subsection{\textsc{Phantom-Geo}}
To determine the statistics, we set up our models similar to \citet{Manzaneda2024}. We set the positions and velocities of the two stars using 
\begin{align}
    x_{i,j} &= x_{\mathrm{cm}} \pm \frac{m_{j,i}}{m_b} \;a\;\cos\varphi\;,\nonumber\\ 
     y_{i,j} &= y_{\mathrm{cm}} \pm \frac{m_{j,i}}{m_b} \;a\;\sin\varphi\;\;,\nonumber\\
    v_{x,i,j} &= v_\mathrm{x,cm} \mp \frac{m_{j,i}}{m_\mathrm{b}} \; a \; \dot{\varphi}\; \sin \varphi\;, \nonumber\\
     v_{y,i,j} &= v_\mathrm{y,cm} \mp \frac{m_{j,i}}{m_\mathrm{b}} \; a \; \dot{\varphi}\; \cos \varphi\;,
\end{align}
where the centre of mass position and velocity is determined by starting single particle orbits with a range of $\beta$ values at a distance of $500\mathord,000\; G M_\bullet/c^2$. The Newton-Raphson method was used to determine the position and velocities of the binary's centre of mass at $50\;r_\mathrm{t}$. We find the pericentre of the geodesic orbit, and use it to determine the correct $\beta$ values. For a retrograde orbit we use
\begin{equation}
   \dot{\varphi} = - \sqrt{\frac{m_\mathrm{b}}{a^3}}\;.
\end{equation}

\subsection{\textsc{Multistar} and \textsc{Tsunami}}
We found that energy equation in \citet{Kidder1995} does not conserve energy. To find the zero-energy orbit, we instead use bisection to determine the orbit, and calculate the position and velocity at $50\;r_\mathrm{t}$. \citet{Manzaneda2024} gives a formula to determine the pericentre velocity for the $\mathrm{PN}$ models (Equation~D15), but we found that this formula gives velocities about $0.1\%$ higher than the ones calculated using our method. This is because the Equation given in \citet{Manzaneda2024} uses Schwarzschild coordinates.

The obtained quantities are used as the centre of mass position and velocity of the binary. We use the \textsc{Multistar} zero-energy orbits positions and velocities obtained at $50\;r_\mathrm{t}$ in \textsc{Tsunami}. We draw the mean anomaly from a equally spaced grid. All orbits are retrograde. 

\begin{figure}
    \centering
    \includegraphics[width=0.6\columnwidth]{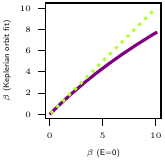}
    \caption{$\beta$ of the orbit calculated using Keplerian orbital parameters compared with \textsc{Multistar} E=0 $\beta$ (purple line). Green line shows one-on-one mapping, but as $\beta \geq 2$, the Keplerian orbital $\beta$ diverges from the actual $\beta$ of the setup. }
    \label{fig:beta_dis}
\end{figure}

To understand how the $\beta$ changes if we use the actual orbital pericentre vs the pericentre calculated from the Keplerian orbit, we determine the pericentre at a long time, where binary is not affected by the black hole's GR effects using
\begin{align}
    \mathbf{e} &= \frac{\mathbf{v}\times\mathbf{h}}{G M_\bullet} - \frac{\mathbf{r}}{r}\;,\nonumber\\
    r_\mathrm{p} &= \frac{h^2}{G M_\bullet (1+e)}\;,
\end{align}
where $\mathbf{h}=\mathbf{r}\times\mathbf{v}$. Figure~\ref{fig:beta_dis} shows that for $\beta \geq 2$, the Keplerian orbital calculation would given lower $\beta$ value. 

\section{EIH Equation implementation in \textsc{Multistar}}
\label{app:eih_multi}
\textsc{Multistar} uses forces along edges between the bodies and reconstructs Jacobi coordinates from these, whereas the \textsc{EIH} formulation in Section~\ref{sec:EIH} is in terms of absolute acceleration.  To implement the equation in the \textsc{EIH} code, we first calculated the absolute accelerations of each particle, and used the \textsc{EIH} equation.

\bibliographystyle{paslike}
\bibliography{bibtemplate}

%\appendix

\end{document}